\documentclass[reprint,superscriptaddress,amsmath,amssymb,aps,pra,twocolumn]{revtex4-2}

\usepackage{subfigure}
\usepackage{tikz}
\usetikzlibrary{quantikz}
\usepackage[colorlinks,linkcolor=blue,anchorcolor=blue,citecolor=blue]{hyperref}
\usepackage{longtable,array}
\usepackage{amsthm,bm,amsmath,multirow}
\newtheorem{theorem}{Theorem}
\newtheorem{lemma}{Lemma}
\newtheorem{definition}{Definition}

\usepackage{xspace}

\hypersetup{%
    plainpages=true,
    breaklinks=true,
    hypertexnames=false,
    pageanchor=true,
    colorlinks=true,
    linkcolor={blue},
    citecolor={blue},
    urlcolor={red},
    anchorcolor={black}
}

\iffalse
\newcommand{\REVISE}[1]{\color{red!80!black}{#1\xspace}\color{black}}
\else
\newcommand{\REVISE}[1]{#1\xspace}
\fi

\newcommand{\casql}{Laboratory of Quantum Information, School of Physics, University of Science and Technology of China, Hefei, Anhui, 230026, P. R. China}
\newcommand{\casex}{\REVISE{Anhui Province Key Laboratory of Quantum Network, University of Science and Technology of China, Hefei, Anhui, 230026, P. R. China}}
\newcommand{\aihf}{Institute of Artificial Intelligence, Hefei Comprehensive National Science Center, Hefei, Anhui, 230088, P. R. China}
\newcommand{\hti}{National Innovation Institute of Defense Technology, AMS, Beijing, 100071, P. R. China}

\newcommand{\origin}{\REVISE{Origin Quantum Computing Technology (Hefei) Co., Ltd., Hefei, Anhui, 230026, P. R. China}}
\newcommand{\iat}{Institute of Advanced Technology, University of Science and Technology of China, Hefei, Anhui, 230031, P. R. China}

\begin{document}
\title{Data-driven Quantum Dynamical Embedding Method for Long-term Prediction on Near-term Quantum Computers}

\author{Tai-Ping Sun}
\affiliation{\casql}
\affiliation{\casex}

\author{Zhao-Yun Chen}
\email{chenzhaoyun@iai.ustc.edu.cn}
\affiliation{\aihf}

\author{Cheng Xue}
\affiliation{\aihf}

\author{Huan-Yu Liu}
\affiliation{\aihf}

\author{Xi-Ning Zhuang}
\affiliation{\casql}
\affiliation{\casex}
\affiliation{\origin}

\author{Yun-Jie Wang}
\affiliation{\iat}

\author{Shi-Xin Ma}
\affiliation{\hti}

\author{Hai-Feng Zhang}
\affiliation{\casql}
\affiliation{\casex}

\author{Yu-Chun Wu}
\email{wuyuchun@ustc.edu.cn}
\affiliation{\casql}
\affiliation{\casex}
\affiliation{\aihf}

\author{Guo-Ping Guo}
\affiliation{\casql}
\affiliation{\casex}
\affiliation{\aihf}
\affiliation{\origin}



\newcommand{\vecm}{\boldsymbol{m}}
\newcommand{\vecx}{\boldsymbol{x}}
\newcommand{\vectheta}{\boldsymbol{\theta}}
\newcommand{\QDE}{\operatorname{QDE}}

\begin{abstract}

     The increasing focus on long-term time series prediction across various fields has been significantly strengthened by advancements in quantum computation. 
     In this paper, we introduce a data-driven method designed for time series prediction with quantum dynamical embedding (QDE). 
     This approach enables a trainable embedding of the data space into an extended state space, allowing for the recursive retrieval of time series information. 
     Based on its independence of time series length, this method achieves depth-efficient quantum circuits that are crucial for near-term quantum computers. 
     Numerical simulations demonstrate the model’s capability to predict not only wave signals but also more complex signals such as NARMA. 
     Prediction accuracy improves with model scaling, and notably, the model achieves better accuracy on wave signal tasks with fewer parameters compared to QRC. 
     Additionally, the model shows promising potential for denoising classical noise in wave signals, and when combined with error mitigation techniques for typical quantum noise, it enables reliable long-term prediction of wave signals.
     We implement this model, restricted to 2 qubits, on the Origin ``Wukong" superconducting quantum processor as a simple proof-of-concept on NISQ devices.
     Furthermore, we provide theoretical analysis of the QDE's dynamical properties for the 2-qubit case and discuss its potential universality.
     Overall, this study represents our first step towards leveraging near-term quantum devices for time series forecasting, offering insights into integrating data-driven learning with quantum dynamical embeddings.

\end{abstract}

\maketitle

\section{Introduction}\label{sec: intro}

\begin{figure*}[!th]
    \centering
    \begin{tikzpicture}
    \node[anchor=center] (center) at (0,0) {
    \includegraphics[width=0.95\textwidth]{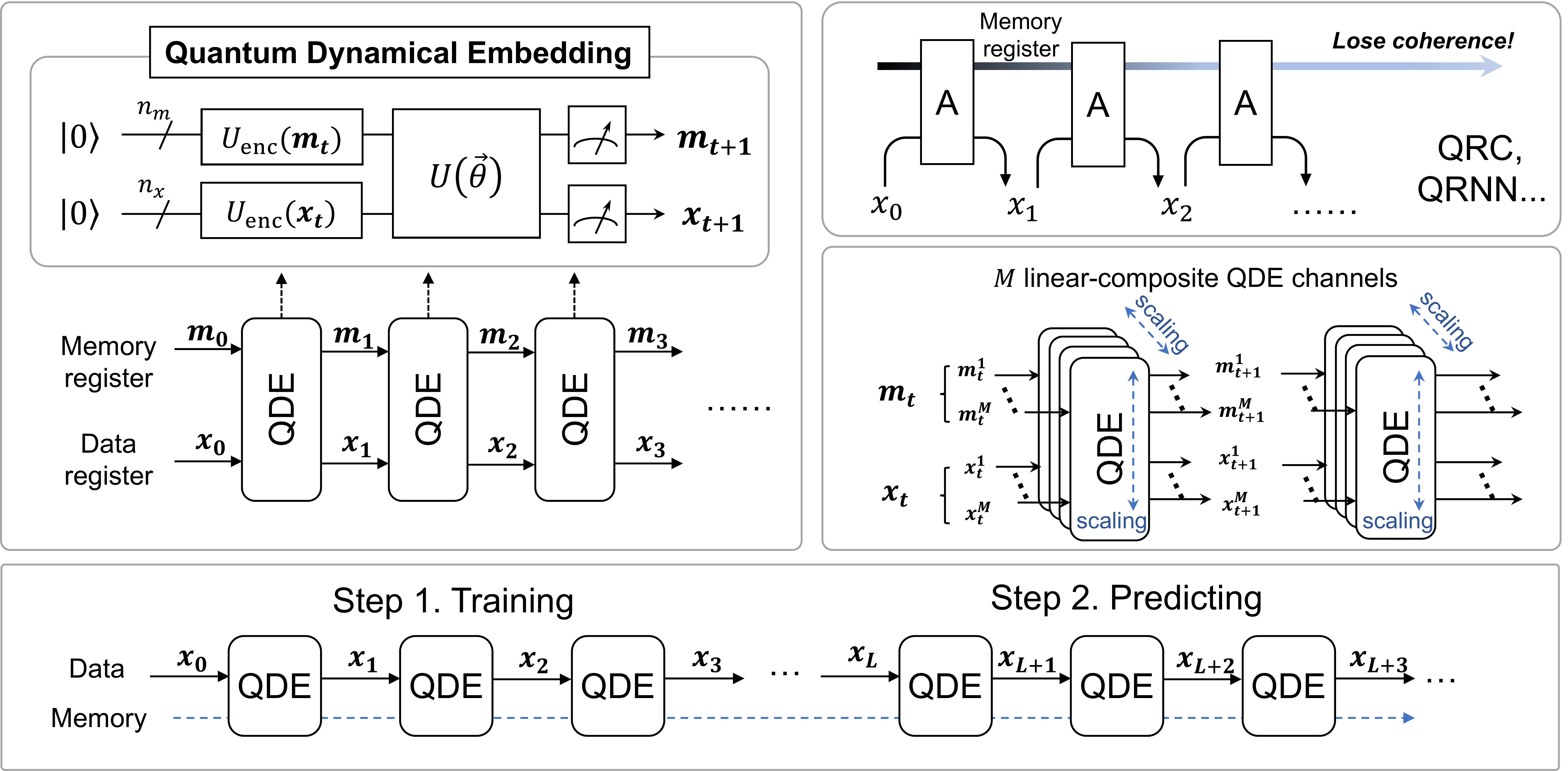} };
    \node[anchor=center] (a) at ($(center.north west)+(0.5cm, -0.5cm)$) {\textbf{(a)}};
    \node[anchor=center] (b) at ($(a.east)+(8.6cm, .0cm)$) {\textbf{(b)}};
    \node[anchor=center] (c) at ($(b.south)+(.0cm, -2.4cm)$) {\textbf{(c)}};
    \node[anchor=center] (d) at ($(a.south)+(0.cm, -5.9cm)$) {\textbf{(d)}};
    \end{tikzpicture}
    \caption{\textbf{Circuits for QDE and Learning Protocols.} (a) Illustration of a generic QDE architecture. The qubits for encoding the classical state are divided into two sets: the memory register (top, with $n_m$ qubits) for inputting memory $\vecm_t$, and the data register (bottom, with $n_x$ qubits) for inputting data $\vecx_t$ at time $t$. 
    The circuit then undergoes evolution through a unitary operator $U(\vectheta)$ under a specified ansatz.
    Finally, selected observable ensembles are measured to extract classical information. 
    With this architecture and the initial point $(\vecm_0, \vecx_0)$, the state pairs $(\vecm_1,\vecx_1),(\vecm_2,\vecx_2),(\vecm_3,\vecx_3),\dots$ are obtained in an autoregressive way. 
    (b) The protocol of QRC and QRNNs. The data is injected into the quantum system and measured to retrieve classical information. 
    The memory register maintains the quantum state form and will lose coherence in the long run. 
    (c) An enhanced QDE architecture with $M$ linear-composite channels. 
    In this design, the QDE is available to explore the dynamics with higher degrees of freedom by either scaling the register qubit number or the composite channel number. 
    The final signal $\vecx_t$ is retrieved through the superposition of sub-signals $\vecx_t^1, \dots, \vecx_t^M$ from different channels. 
    (d) Training and predicting protocols. 
    In the training stage, initial data $\vecx_0$ is fed into the QDE block, and the output data $\hat{\vecx}_{i}$ is used to compute the cost function, with $i= {1,\cdots,L}$ and time updated over $L$ steps. 
    After the training, the system undergoes an additional $T$ steps of evolution to predict future values. 
    Note that while the memory evolves simultaneously with the data in both stages, it does not contribute to the cost function calculation. 
    }
    \label{fig: illustr}  
\end{figure*}

Recent years have witnessed a surge in the application of time series prediction across a range of fields, including finance, physics, and engineering, underscoring its significance~\cite{Shumway2000, Kantz2004, Winterhalder2006}.
Among the various types of predictions, long-term prediction is especially pivotal as it entails forecasting future values of a series that extend beyond the duration of the available historical data. 
This means that the prediction horizon is longer than the time span of the training dataset, presenting a unique challenge in identifying and extrapolating underlying patterns and trends that may continue to evolve over time~\cite{Gardner1984, Fildes2004}. 
The importance of long-term prediction is evident in strategic planning for climate change, economic forecasting, and technological forecasting, where decisions made today are based on projections that reach far into the future~\cite{Bauer2015, Lam2020}.
Classical algorithms for time series prediction predominantly rely on long short-term memory (LSTM) networks~\cite{Hochreiter1997, Cho2014}, echo state networks (ESNs)~\cite{Jaeger2001, Jaeger2004}, and reservoir computing (RC)~\cite{Nakajima2021book}. 
These data-driven methods have shown potential in capturing dynamical features. 
Recent advances in quantum computing have demonstrated significant potential in the machine-learning domain~\cite{Mitarai2018,Rebentrost2014,Lloyd2018,Cerezo2021REW}. 
Quantum computers offer a promising avenue to overcome classical computational bottlenecks by enabling coherent processing of data within exponentially large Hilbert spaces, providing potential quantum advantages. 
Therefore, it is both intriguing and important to explore quantum-based approaches focused on time series processing, especially for realistic and challenging tasks such as long-term time series prediction.
Recently, it has been demonstrated that classical data can be mapped into a quantum feature space using certain embeddings~\cite{Rebentrost2014, Lloyd2018, Havlicek2019, Huang2021Power}, which is further utilized to approximate a variety of functions.
These universal approximation properties~\cite{Mitarai2018, Schuld2021, Goto2021, Yu2022} facilitate the construction of inference functions for modeling temporal data patterns, suggesting the potential of quantum systems to effectively model the complex dynamics inherent in long-term time series.

Current quantum algorithms for time series processing, such as quantum reservoir computing (QRC)~\cite{Fujii2017, Nakajima2019, Chen2020, Fujii2021, Martinez2021, Mujal2021, Tran2021, Govia2022, Kalfus2022, Pfeffer2022, Suzuki2022, Milan2024, Monomi2025}, quantum recurrent neural networks (QRNNs)~\cite{Bausch2020, Takaki2021, Chen2022}, and quantum dynamic mode decomposition (QDMD)~\cite{Xue2023}, have demonstrated significant potential in understanding temporal dependencies in sequential data. 
However, in QRC and QRNNs, the circuit depth is determined by a constant factor ($C$) from the ansatz, multiplied by the time sequence length ($L$), i.e., $C \times L$, which poses challenges for noisy intermediate-scale quantum (NISQ) devices.
The increasing circuit depth threatens the maintenance of quantum coherence over time, which is crucial for the accuracy of long-term predictions. 
Although the QRC model has shown significant performance of time series prediction, its requirement for deep circuits to maintain temporal dependencies makes it impractical for NISQ devices due to exacerbated decoherence and noise, which undermine the performance of quantum algorithms in realistic settings.
Efficiently utilizing these devices for complex tasks like long-term time series prediction without deep circuits is essential for advancing practical quantum computing applications. 
Therefore, there is a pressing need for methods that can predict long-term time series with shallow quantum circuits, addressing the limitations posed by current quantum hardware. 
Solutions have been explored through both theoretical and physical implementations on various platforms~\cite{Negoro2018, Negoro2021, Nokkala2021, Mujal2023, Dudas2023, Garcia-Beni2024, Yasuda2023}, yet it remains an open question.

In this study, we present a data-driven approach for time series prediction using quantum dynamical embedding (QDE). Our approach allows for the dynamic embedding of original data into an extended state space and facilitates the data-driven learning of embedding \REVISE{via trainable quantum-gate parameters, i.e., in a quantum-neural-network style as in~\cite{Takaki2021}. It differs from feedback-based methods~\cite{Kobayashi2024,Ehlerrs2025,Monomi2025} in the placement and optimization of tunable parameters: the parameters reside within the circuit ansatz and optimized iteratively, rather than learned by regression at the readout as in~\cite{Fujii2017}}. 
Specifically, we recurrently retrieve time series data from measurements of the QDE at the current time step.
This design splits the long-term time sequence into independent input-prediction pairs, ensuring that for each prediction task the qubit count and circuit depth $C$ are determined by the ansatz architecture, irrespective of the time series length. Although the circuit depth is reduced, a full prediction of the time sequence is achieved by repeating the circuit $L$ times with different inputs. Consequently, it mitigates issues related to decoherence and error accumulation over time.

Embedding serves as an effective strategy for converting from non-Markovian to Markovian dynamics by expanding the hidden space, allowing for a more flexible representation of temporal data. 
This enlargement is pivotal for revealing the intricate dynamics and temporal patterns inherent in the series. 
Furthermore, the flexibility to choose the size of the extended hidden space and the trainability of the embedding parameters are particularly advantageous for long-term time series prediction, ensuring a data-driven approach that adapts to the complexity of the data at hand.

We demonstrate the effectiveness of our method through numerical experiments on wave signal prediction as well as more complex tasks like NARMA.
Additionally, we evaluate its denoising capabilities on the same prediction tasks when exposed to random noise, further showcasing the QDE's versatility.
Notably, a comparative analysis with existing methodologies reveals that our approach offers significant advantages in terms of average prediction accuracy and computational efficiency on the same tasks. 
We also recognize limitations related to the training procedure, such as the presence of barren plateau, local minima and instability, which point to important directions for further model optimization.

To validate the noise resistance of our method, we perform long-term predictions using a noisy simulator with a learnable error-cancellation layer (LECL) following the QDE. Additionally, we extend our implementation to the Origin ``Wukong'' superconducting quantum computer, integrating an LECL to generate time series for proof-of-concept demonstration. This demonstrates the potential of leveraging current quantum computing capabilities for real-world time series prediction challenges.

Theoretically, we explore the dynamics of a specific 2-qubit case, providing insights into the connection between the QDE and the inherent characteristics of time series. 
Developing a more general theoretical framework, together with demonstrating performance improvements for broader time series processing tasks, remains an open challenge.
Although this approach does not fully replace QRC, the data embedding method offers a promising and insightful direction for quantum machine learning applied to time series analysis.


\section{Method}\label{sec: method}

\subsection{The quantum dynamical embedding model} \label{sec: qde}

The common target among different time series prediction tasks can be viewed as learning the complex system's dynamics from a huge size of historical data.
Markovian dynamics can be characterized by the iteration, $\vecx_{t+1}=f(\vecx_t)$,
where $f(\cdot)$ represents the update function, and $\vecx_t$, $\vecx_{t+1}$ represent the system state at time steps $t$ and $t+1$, respectively. In more complex non-Markovian scenarios, a memory architecture is necessary to maintain the autonomous nature of the system: $(\vecx_{t+1}, \vecm_{t+1}) = \mathcal{F}(\vecx_{t}, \vecm_{t})$, where $\mathcal{F}(\cdot)$ denotes the extended update function, $\vecm_{t}$ and $\vecm_{t+1}$ denote the hidden states at time steps $t$ and $t+1$, respectively.
Such an embedding, from the low-dimensional yet non-Markovian space of original data into a high-dimensional Markovian phase space, has shown great power in RC, ESNs, and recurrent neural networks~\cite{Nakajima2021book, Jaeger2004, Rumelhart1986, Elman1990, Jordan1997, Hochreiter1997, Cho2014}.

However, this embedding procedure turns out to be too expansive when training a highly non-Markovian process with a quickly growing phase space.
Fortunately, quantum systems, due to their exponentially large Hilbert space, can be leveraged as a work memory to encode and evolve the phase space~\cite{Negoro2018, Negoro2021, Mujal2023, Takaki2021}.
Nevertheless, there are still two essential weaknesses faced by all of the works mentioned above: 
Firstly, a long-term coherence needs to be preserved which seems formidable in the NISQ era.
Secondly, the circuit depth grows at least linearly in those approaches, inducing a strong restriction on any further long-term training or predicting procedures.

To address this issue, we propose a classical-quantum hybrid protocol to learn and predict long-term time series, wherein only fixed-depth circuits are employed, and long-term training and predicting are enabled, as shown in Fig.~\ref{fig: illustr}.
More specifically, for each step, given the hidden memory $\vecm_t$ of size $n_m$ and the observable data $\vecx_t$ of size $n_x$,
a shallow parameterized quantum circuit $(n_m, n_x)$-QDE is applied to the input pair $(\vecm_t, \vecx_t)$
\begin{equation}
    |\psi_t\rangle = U(\vectheta)|\vecm_t\rangle_m|\vecx_t\rangle_x,
\end{equation}
where 
$$|\vecm_t\rangle_m|\vecx_t\rangle_x = \left(U_{\text{enc}}(\vecm_t)|0\rangle_{m}\right)\otimes\left(U_{\text{enc}}(\vecx_t)|0\rangle_{x}\right)$$
is the encoded input state. $U(\vectheta)$ is a trainable parameterized quantum circuit. \REVISE{The encoding strategies $U_{\text{enc}}$ may vary; a commonly used scheme is introduced later in Sec.~\ref{sec: settings}. Our method relies on the assumption that the embedding effectively maps the original non-Markovian dynamics to an (approximate) Markovian form. Accordingly, in the present work, we neglect time-delayed memory terms, denoted $\bm{m}_{t-i}$ for $i>0$. } The memory and data of the next step are extracted from observing of their corresponding subspaces, namely
\begin{equation} \label{eq: defm}
    \vecm_{t+1} = \operatorname{Tr}\left(\mathcal{H}_{m}|\psi_t\rangle\langle\psi_t|\right),
\end{equation}
and
\begin{equation} \label{eq: defx}
    \vecx_{t+1} = \operatorname{Tr}\left(\mathcal{H}_{x}|\psi_t\rangle\langle\psi_t|\right),
\end{equation}
where $\mathcal{H}_{m}$ and $\mathcal{H}_{x}$ are the observable vector with length $n_m$ and $n_x$, respectively. In the following, we explicitly consider the Pauli-$\sigma_z$ operator as the observable. This means that $(\mathcal{H}_m,\mathcal{H}_x)$ correspond to measuring $\sigma_z$ on each qubit individually.

This process gives a mapping from $(\vecm_t, \vecx_t)$ to $(\vecm_{t+1}, \vecx_{t+1})$, which is denoted as $\QDE(\vecm_t, \vecx_t ;\theta) \mapsto (\vecm_{t+1}, \vecx_{t+1})$. Given any parametrized quantum circuit $U(\vectheta)$ and the initial state $(\vecm_0, \vecx_0)$, one could keep performing this mapping to obtain $\vecx_1$, $\vecx_2$, ... $\vecx_{\infty}$, as illustrated in Fig.~\ref{fig: illustr}(a). 

The QDE distinguishes itself from previous studies of QRC and QRNNs in the manner of memory information storage. Unlike these models, where the quantum system serves as a memory reservoir for historical data, our approach overcomes the issue of coherence loss---a challenge in long-term data embedding and retrieval---through the QDE model, as illustrated in Fig.~\ref{fig: illustr}(b). \REVISE{Recent works~\cite{Kobayashi2024,Monomi2025} also adopt a reinjection-based approach to memory, but differ from QDE in their memory-state reconstruction procedures and in the overall learning and optimization framework.}

Enhancements to the QDE model are achieved by integrating $M$ linear-composite channels, as shown in Fig.~\ref{fig: illustr}(c). Each channel in the QDE utilizes the architecture in Fig.~\ref{fig: illustr}(a), albeit with distinct parameters. These channels are designed to contain identical qubit numbers for both data and memory registers. The final output, $\vecx_t$, is a synthesis of sub-signals $\vecx_t^1,\dots,\vecx_t^M$ from these varied channels, enabling the QDE to explore dynamics with increased degrees of freedom.

\subsection{Training and predicting} \label{sec: protocols}
As we can see from the QDE model, with a fixed circuit $U(\vectheta)$ and the initial point, we can determine the evolution of this system. Given the trainability of the parametrized quantum circuit, we can employ a data-driven method for predicting the time series. That is, we utilize historical data points to develop a model that accurately represents the underlying patterns in the data. Once this model is established, we extrapolate it over time to forecast future trends and values. The training of the model can be formulated as
\begin{definition}[Training a data-driven model]
    Given a trainable predictive model $\mathcal{F}(\vecx, \vecm;\vectheta)$, where $\vecx$ denotes the data, $\vecm$ the hidden memory, $\vectheta$ the trainable parameters, and a standard historical dataset $\vecx_{0:L}$, where $\vecx_{0:L}$ represents the series of data points from time 0 to L. Use this model to sequentially generate a predicted data $\hat{\vecx}_{0:L}$ from the initial point $\vecx_{0}$. The objective is to optimize the parameter $\vectheta$, formulated as
    $$\operatorname{argmin}_{\vectheta} \mathcal{L}(\hat{\vecx}_{0:L},\vecx_{0:L};\vectheta)$$ where $\mathcal{L}$ quantifies the difference between the predicted and historical series.
\end{definition}

In the subsequent applications, the QDE is optimized using mean square error (MSE) as the loss function, defined as:
\begin{equation}\label{eq:mse}
    \mathcal{L}_{\mathrm{mse}}=\frac{1}{L}\sum_{t=1}^L\left\|\vecx_t-\hat{\vecx}_t\right\|_2^2,
\end{equation}
where the $\hat{\vecx}_t$ and $\vecx_t$ denote the time series values at time step $t$ for $\hat{\vecx}_{0:L}$ and $\vecx_{0:L}$, respectively. The term $\|\cdot\|_2$ represents the $L^2$ norm. 

Building on the findings of previous research~\cite{Fujii2017,Takaki2021}, the embedding of time series and the evolving of quantum systems require a circuit depth of $\mathcal{O}(L)$. Comparatively, our approach marks a notable advancement by achieving this with a significantly reduced circuit depth of $\mathcal{O}(\Gamma)$, where $\Gamma$ represents the base depth of a single QDE circuit block.

Moreover, we employ gradient descent for circuit parameter optimization, with the number of circuit executions scaling as $\mathcal{O}(nL)$, where $n=n_m+n_x$ denotes the total qubit number in the QDE. The linear dependency on the time series length $L$ facilitates the use of gradient-based optimization methods. For an in-depth discussion of this procedure, see Sec.~\ref{sec: gradient}.

After training, the iterative mapping is applied over $T$ additional steps for prediction, generating the series $\hat{\vecx}_{L+1:L+T}$ to approximate target values. In the subsequent tests, we compare the prediction results with the ideal results to show the performance of the QDE. The predicting error is also defined as MSE.

\subsection{Gradient evaluation}\label{sec: gradient}
The optimal parameters for the QDE are obtained by minimizing the loss function, as expressed in Eq.~\eqref{eq:mse}. During the training stage, the initial state vector $(\vecm_0, \vecx_0)$ at time $t=0$ is composed of the given initial data $\vecx_0$ and its complementary component $\vecm_0$, which is initialized randomly. This state vector is then fed into the QDE block, yielding a  time sequence $\{(\vecm_1, \hat{\vecx}_1), \dots, (\vecm_L, \hat{\vecx}_L)\}$, with the explicit part represented as $\{\hat{\vecx}_1, \dots, \hat{\vecx}_L\}$. 
To simplify the illustration, we focus on a 2-qubit QDE, where both the given data and the memory are scalars. The loss function is then reformulated as $\mathcal{L}_{mse}=\frac{1}{L}\sum_{t=1}^{L}\left(\hat{x}_t-x_t\right)^2$. The derivative of $\mathcal{L}_{mse}$ with respect to $\theta_i$ is given by:

\begin{align}\label{eq:grad}
    \frac{\partial \mathcal{L}_{mse}}{\partial \theta_i} &= \frac{2}{L}\sum_{t=1}^L \left(\hat{x}_t-x_t\right)\frac{\partial \hat{x}_t}{\partial \theta_i}.
\end{align}

Referring to the definition of $\hat{x}_{t+1}$ in Eq.~\eqref{eq: defx} and the explicit expression of $\mathcal{H}_{m}=\sigma_z^0$ and $\mathcal{H}_{x}=\sigma_z^1$, this expression can be equivalently written as:
\begin{equation}\label{eq:grad1}
    \begin{aligned}
        \frac{\partial \hat{x}_{t+1}}{\partial \theta_i} &= \frac{\partial }{\partial \theta_i}\operatorname{Tr}\left(\sigma_z^1|\psi_t\rangle\langle\psi_t|\right). \\
        &= \frac{\partial }{\partial \theta_i}\mathrm{Tr}\left[\sigma_z^1 U(\bm{\theta})U_{enc}(\hat{X}_t)\rho_0 U_{enc}^\dagger(\hat{X}_t)U^\dagger(\bm{\theta})\right],
    \end{aligned}
\end{equation}

where $\hat{X}_t=(m_t,\hat{x}_t)$ and $\rho_0 =\left(|0\rangle\langle 0|\right)^{\otimes n}$ for simplicity.
It is important to note that both $U(\bm{\theta})$ and $U_{enc}(\hat{X}_t)$ depend on $\theta_i$. Hence, the derivative of Eq.~(\ref{eq:grad1}) is:

\begin{widetext}
\begin{equation}\label{eq:grad2}
    \begin{aligned}
        \frac{\partial \hat{x}_{t+1}}{\partial \theta_i} &= \mathrm{Tr}\left[\sigma_z^1 \frac{\partial U(\bm{\theta})}{\partial \theta_i}U_{enc}(\hat{X}_t)\rho_0U_{enc}^\dagger(\hat{X}_t)U^\dagger(\bm{\theta})\right]+ \mathrm{Tr}\left[\sigma_z^1 U(\bm{\theta})U_{enc}(\hat{X}_t)\rho_0U_{enc}^\dagger(\hat{X}_t)\frac{\partial U^\dagger(\bm{\theta})}{\partial \theta_i}\right] \\
        &\quad+ \mathrm{Tr}\left[\sigma_z^1 U(\bm{\theta})\frac{\partial U_{enc}(\hat{X}_t)}{\partial \theta_i}\rho_0U_{enc}^\dagger(\hat{X}_t)U^\dagger(\bm{\theta})\right] + \mathrm{Tr}\left[\sigma_z^1 U(\bm{\theta})U_{enc}(\hat{X}_t)\rho_0\frac{\partial U_{enc}^\dagger(\hat{X}_t)}{\partial \theta_i}U^\dagger(\bm{\theta})\right].
    \end{aligned}
\end{equation}
\end{widetext}

The first two terms on the right-hand side of Eq.~(\ref{eq:grad2}) can be computed using the parameter shift rule~\cite{Schuld2019}, as the rotation gates are generated with Pauli operators, e.g., $R_Y(\theta_i) = e^{-i\theta_i/2\sigma_{y}}$. Assuming that the encoding gates are also Pauli-generated unitaries and that the function $u(x) = \arccos(x)$ is designed to convert state to rotation angle, the remaining terms on the right-hand side of Eq.~(\ref{eq:grad2}) can be computed as follows:

\begin{widetext}
\begin{equation}\label{eq:grad3}
    \begin{aligned}
        \frac{\partial U_{enc}(\hat{X}_t)}{\partial \theta_i} &= \frac{\partial U_{enc}(m_t)}{\partial \theta_i}\otimes U_{enc}(\hat{x}_t) + U_{enc}(m_t)\otimes \frac{\partial U_{enc}(\hat{x}_t)}{\partial \theta_i}\\
        &= \frac{\partial U(u)}{\partial u}\frac{\partial u(m_t)}{\partial m_t}\frac{\partial m_t}{\partial \theta_i}\otimes U_{enc}(\hat{x}_t) + U_{enc}(m_t)\otimes \frac{\partial U(u)}{\partial u}\frac{\partial u(\hat{x}_t)}{\partial \hat{x}_t}\frac{\partial \hat{x}_t}{\partial \theta_i}\\
        &= \left(-\frac{1}{\sqrt{1-m_t^2}}\cdot \frac{\partial m_t}{\partial \theta_i} \right)\frac{\partial U(u)}{\partial u}\otimes U_{enc}(\hat{x}_t) + \left(-\frac{1}{\sqrt{1-\hat{x}_t^2}}\cdot \frac{\partial \hat{x}_t}{\partial \theta_i} \right)U_{enc}(m_t)\otimes \frac{\partial U(u)}{\partial u}.
    \end{aligned}
\end{equation}
\end{widetext}
Together with its conjugate term $\partial U_{enc}^\dagger(\hat{X}_t)/\partial \theta_i$, each term in the last equation of Eq.~(\ref{eq:grad3}) can be efficiently calculated using the parameter shift rule. This is because the coefficients can be evaluated with the current values of $m_t$, $\hat{x}_t$, and previous values of $\partial m_t/\partial \theta_i$, $\partial \hat{x}_t/\partial \theta_i$. 
Thus, while the gradient of $\hat{x}_{t+1}$ with respect to variable $\theta_i$ depends on all previous time steps, the gradient for a single time step can be computed in 6 evaluations based on the parameter shift rule, provided that the partial derivatives $\partial m_t/\partial \theta_i$, $\partial \hat{x}_t/\partial \theta_i$ are computed in order $t=1,2,\dots,L$. Since $m_{t+1}$ is obtained with the same quantum circuit as $\hat{x}_t$ but with a different measurement $\mathcal{H}_m$, computing $\partial m_t/\partial \theta_i $ will not increase the number of evaluations. 
Taking into account the sampling error for each expectation value $\langle\sigma_z\rangle$, the total sample complexity for estimating the gradient $\partial \mathcal{L}_{mse}/\partial \bm{\theta}$ scales as $\mathcal{O}(\frac{4(n+1)L | \{\theta_i\}|}{\varepsilon^2})$, where $n$ is the number of qubits, $L$ is the sequence length, $| \{\theta_i\}|$ is the number of circuit parameters, and $\varepsilon$ denotes the desired precision for observable measurements.

As a specialized type of quantum neural network (QNN), it is natural to consider whether the QDE model suffers from the barren plateau phenomenon, where the loss function or its gradients concentrate and vanish as the system size increases. This issue has been extensively studied, with analyses of its origins and methods to mitigate or avoid it~\cite{McClean2018,Cerezo2021,Marrero2021,Uvarov2021,Liu2023,Larocca2025}. Since the QDE is constructed from the QNN building blocks, it is important to evaluate the impact of barren plateaus on its trainability. In Section~\ref{sec: applic}, we provide numerical results assessing the gradient concentration of the QDE model to better understand its behavior in this regard.

\subsection{Experimental settings}\label{sec: settings}

The quantum circuit of the QDE comprises three primary components: encoding, parametrized evolution, and quantum measurement, as detailed in Sec.~\ref{sec: qde}. 
To simplify the QDE circuit, we implement the encoding layer $U_{enc}(\cdot)$ with rotation $R^i_Y(\arccos (\cdot))$ for the $i^{\text{th}}$ qubit. 
The input is normalized to the range $[-1,1]^{\otimes n}$.
This design choice is driven by the understanding that the input can be efficiently reconstructed using the $\sigma_z$ expectation, eliminating the need for additional evolution.
As the number of qubits, circuit depth, and linear-composite channels increase, the QDE exhibits more complex dynamics. 
The methodology for scaling these elements is depicted in Fig.~\ref{fig: illustr}(c) and properties are further discussed in Sec.~\ref{sec: analysis}. 
Throughout the time sequence iterations, all quantum circuit parameters remain constant. 

The QDE's circuit architecture is tailored to match the complexity of the specific task at hand. In our study, we propose two distinct quantum circuit architectures: the transverse Ising evolution ansatz (TIEA) and the hardware-efficient ansatz (HEA).

\textbf{TIEA:} This architecture is designed to enhance the expressiveness of each ansatz layer. The unitary operator $U(\bm{\theta})$ includes sequential single-qubit rotations followed by a Hamiltonian evolution applied across the entire circuit. The single-qubit rotations are defined by the equation:
\begin{equation}\label{eq:singlegate}
U_1(\theta_1, \theta_2, \theta_3) = R_X(\theta_1)R_Z(\theta_2)R_X(\theta_3),
\end{equation}
with $\theta_1$, $\theta_2$, and $\theta_3$ as real parameters from the set $\vectheta$, and $R_X$ and $R_Z$ as single-qubit rotations around the x and z axes, expressed as $R_X(\theta) = e^{-\frac{i}{2}\theta \sigma_x}$ and $R_Z(\theta) = e^{-\frac{i}{2}\theta \sigma_z}$, respectively. These rotations are followed by a Hamiltonian evolution, denoted as $e^{-iH\tau}$, where $\tau$ signifies the evolution time and $H$ the Hamiltonian, specified as in previous works~\cite{Mitarai2018K, Takaki2021, Martinez2021}:
\begin{equation}\label{eq:Hamiltonian}
H = \sum_{i=1}^n (h+D_i) \sigma_{x,i} + \sum_{i=2}^n \sum_{j=1}^{i-1} J_{ij}\sigma_{z,i}\sigma_{z,j},
\end{equation}
where $h$ represents the transverse field strength and $J_{ij}$ denotes the coupling strength of site $i$ and $j$.
The coefficients $D_i, J_{ij}$ are randomly chosen from a uniform distribution within $[-W,W]$ and $[-J_s, J_s]$, respectively. And all  remain fixed during training. 

\textbf{HEA~\cite{Kandala2017}:} 
This architecture can be viewed as a variant of TIEA, where the Hamiltonian evolution layer is replaced with a CZ entanglement layer, while retaining the single-qubit rotation gates $U_1$. For certain time series tasks discussed in the next section, the rotation gate can be further simplified. For certain time series tasks discussed in the next section, the rotation gate can be further simplified.

\section{Results}\label{sec: results}
\begin{figure*}[!t]
    \centering
    \begin{tikzpicture}
        \node[anchor=north east] (img) at (0,0) {\includegraphics[width=1.0\textwidth]{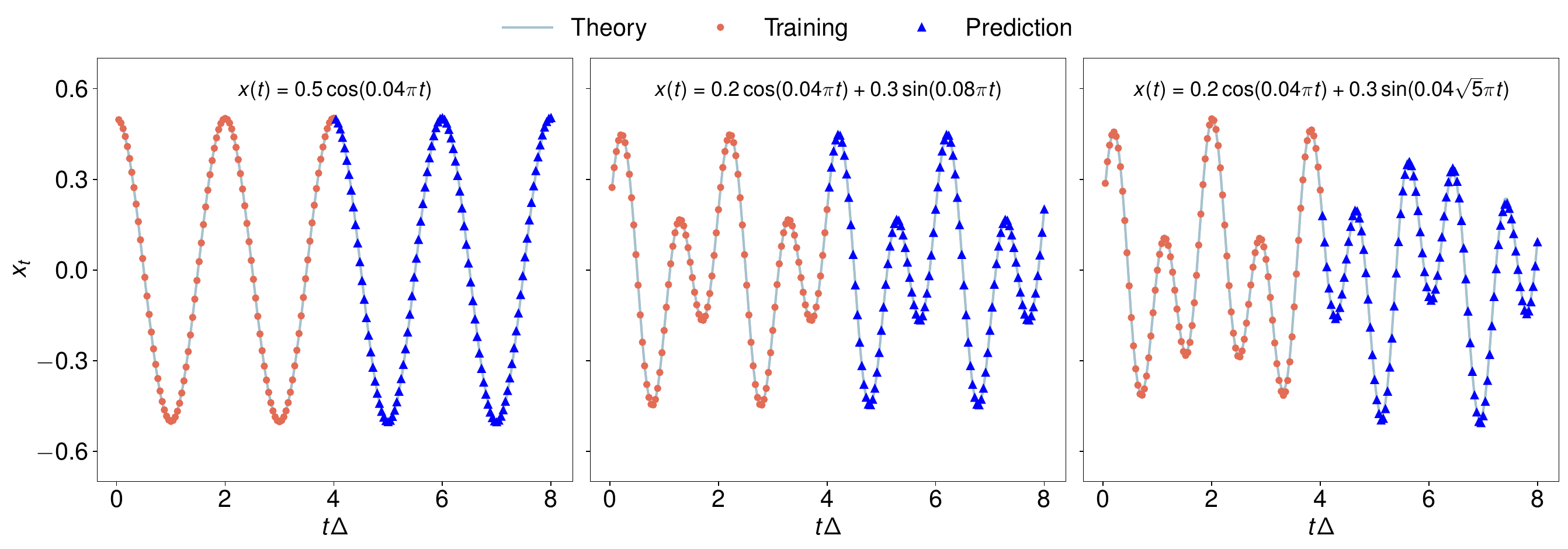}};
        \node[anchor=north east] (img2) at (0, -6.2) {\includegraphics[width=1.0\textwidth]{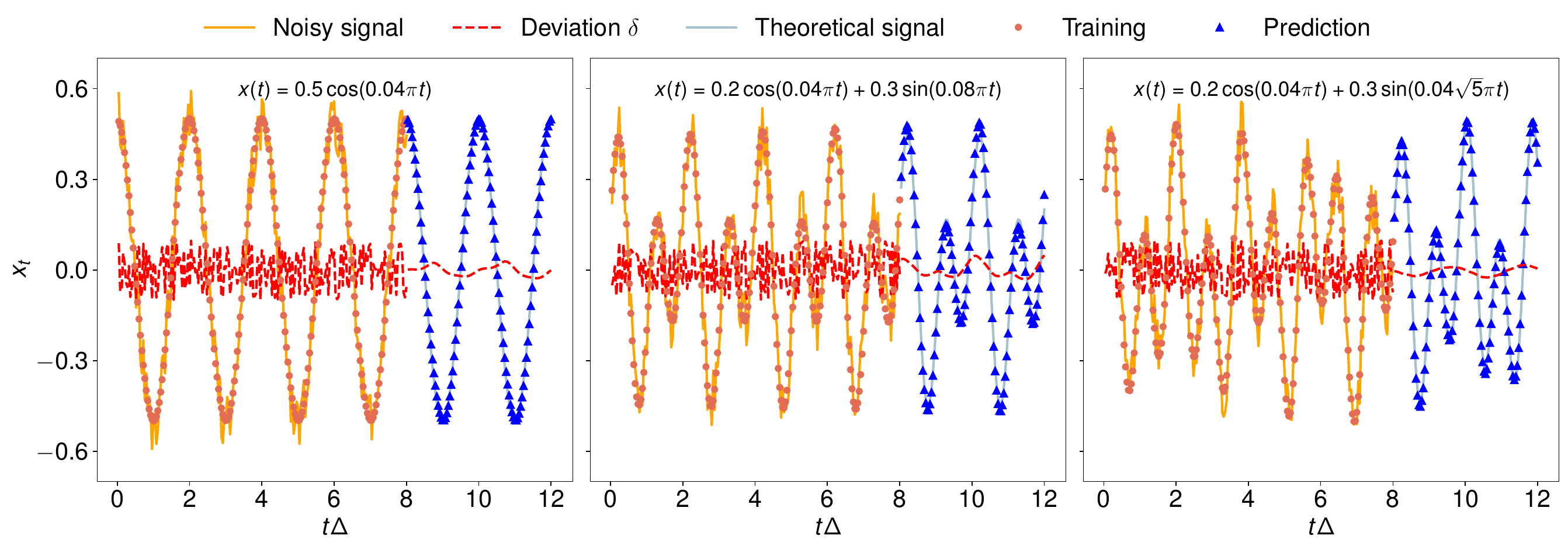}};
        \node[anchor=north east] (img3) at (0, -11.8) {\includegraphics[width=1.0\textwidth]{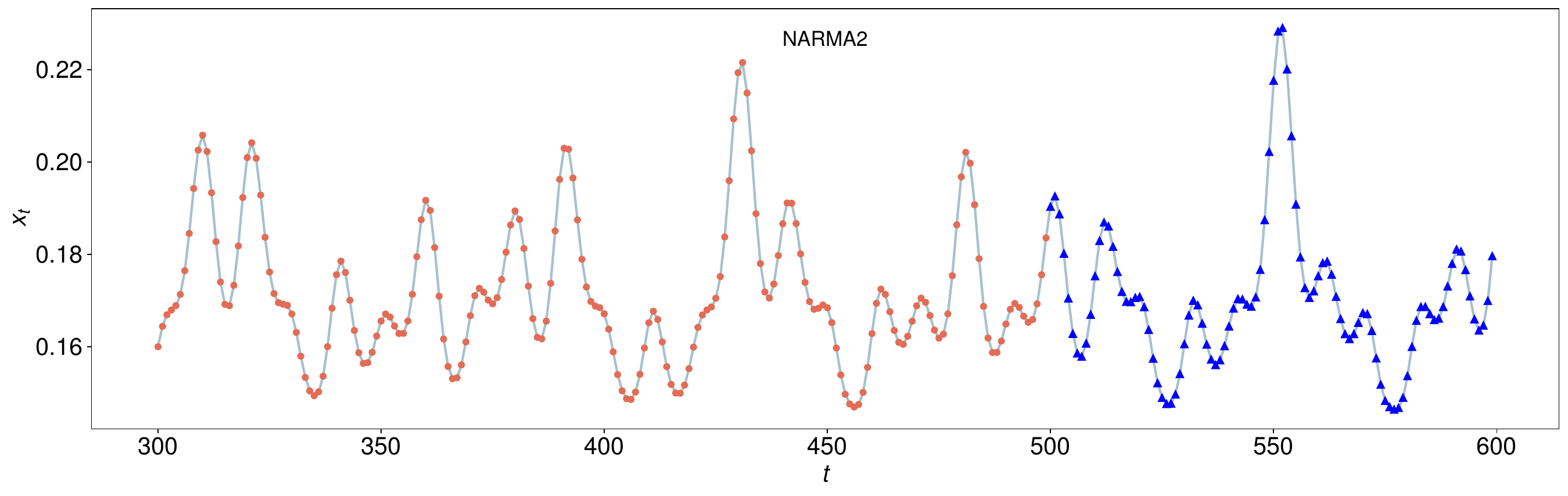}};
        \node[anchor=north east, inner sep=2pt] at ([xshift=-16.3cm, yshift=-0.8cm] img.north east) {\textbf{(a)}};
        \node[anchor=north east, inner sep=2pt] at ([xshift=-10.65cm, yshift=-0.8cm] img.north east) {\textbf{(b)}};
        \node[anchor=north east, inner sep=2pt] at ([xshift=-5.1cm, yshift=-0.8cm] img.north east) {\textbf{(c)}};
        \node[anchor=north east, inner sep=2pt] at ([xshift=-16.3cm, yshift=-0.2cm] img2.north east) {\textbf{(d)}};
        \node[anchor=north east, inner sep=2pt] at ([xshift=-10.65cm, yshift=-0.2cm] img2.north east) {\textbf{(e)}};
        \node[anchor=north east, inner sep=2pt] at ([xshift=-5.1cm, yshift=-0.2cm] img2.north east) {\textbf{(f)}};
        \node[anchor=north east, inner sep=2pt] at ([xshift=-16.3cm, yshift=-0.2cm] img3.north east) {\textbf{(g)}};
    \end{tikzpicture}
    \caption{\textbf{Demonstrations of QDE for Time Series Prediction}: (a) $(1,1)$-QDE application on cosine-wave prediction with $x(t) = 0.5\cos(\omega t)$, where $\omega = \pi/25$ and the value of time axis is rescaled by $\Delta=0.04$. The training and predicting lengths are set to $L=T=100$ in this and the following tests, unless otherwise specified. (b) Composite periodic signal predicting, $x(t) = 0.2\cos(\omega t) + 0.3\sin(2\omega t)$, where $\omega = \pi/25$. (c) Aperiodic time series predicting, $x(t) = 0.2\cos(\omega t) + 0.3\sin(\sqrt{5}\omega t)$, with the same architecture and settings as (b). 
    (d), (e), and (f) correspond to the noisy versions of the scenarios described in (a), (b), and (c), respectively, trained with noise uniformly distributed within $[-0.1, 0.1]$. 
    For these versions, the number of training points is increased to enhance the capture of signal characteristics buried in noise. 
    The prediction part of each subplot is juxtaposed with theoretical clean signals for comparison. 
    The red dashed line represents the deviation $\delta  = \hat{x}(t)-x_{\text{ref}}(t)$, where $\hat{x}(t)$ is the output of the QDE model and $x_{\text{ref}}(t)$ is the reference, the noisy signal in training stage and theoretical signal in predicting stage. 
    (g) Prediction of NARMA2 data using the $(2, 1)$-QDE model. The training length is set to 500 steps, with a fixed prediction length. For clarity, only the last 200 training points are shown. The signal is generated following the formula described in the main text.
    }

    \label{fig: applic}  
\end{figure*}

\subsection{Applications}\label{sec: applic}

\newcommand{\NAline}{\raisebox{0.5ex}{\rule{.5cm}{0.5pt}}}
\newcolumntype{P}[1]{>{\centering\arraybackslash}p{#1}}
\newcommand{\colwidth}{1.3cm}

\begin{figure*}[!t]
    \centering
    \begin{tikzpicture}
        \node[anchor=north east] (img) at (0,0) {\includegraphics[width=1.0\textwidth]{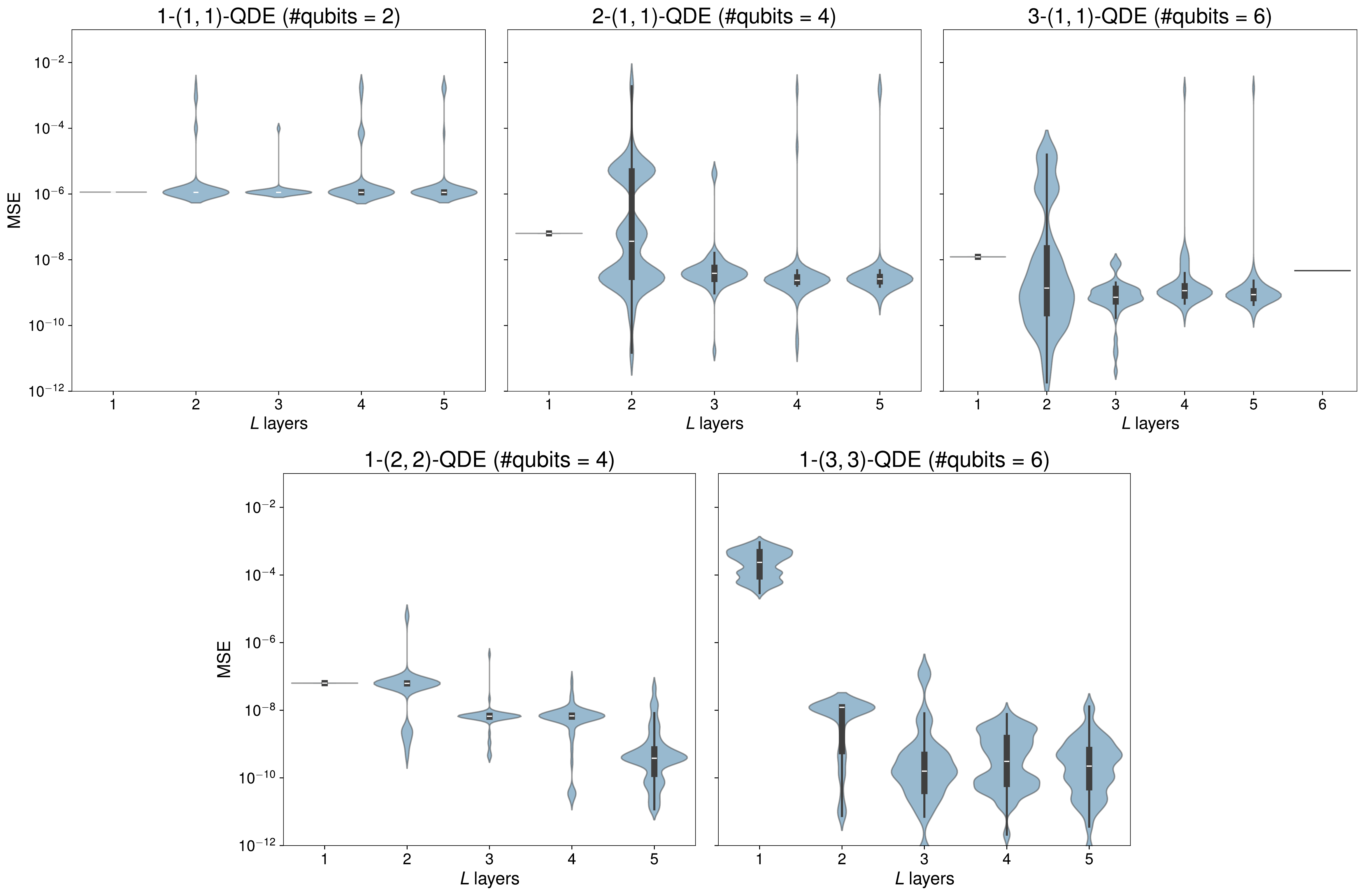}};
        \node[anchor=north east, inner sep=2pt] at ([xshift=-16.5cm, yshift=-0.6cm] img.north east) {\textbf{(a)}};
        \node[anchor=north east, inner sep=2pt] at ([xshift=-10.7cm, yshift=-0.6cm] img.north east) {\textbf{(b)}};
        \node[anchor=north east, inner sep=2pt] at ([xshift=-5.05cm, yshift=-0.6cm] img.north east) {\textbf{(c)}};
        \node[anchor=north east, inner sep=2pt] at ([xshift=-13.7cm, yshift=-0.2cm] img2.north east) {\textbf{(d)}};
        \node[anchor=north east, inner sep=2pt] at ([xshift=-8cm, yshift=-0.2cm] img2.north east) {\textbf{(e)}};
    \end{tikzpicture}
    \caption{\textbf{Performance Scaling of the QDE Model for Cosine-wave Signal Prediction Under Two Scenarios.} (a)-(c) The first row demonstrates the effects of increasing the total number of channels in a $(1,1)$-QDE configuration. (d)-(e) The second row illustrates performance changes when scaling up the number of internal qubits within a single QDE block (i.e., increasing $n_x$ and $n_m$). Each violin plot represents the distribution of MSE outcomes across 50 trials, highlighting median trends and variability associated with different circuit depths and qubit counts.
    }
    \label{fig: scaling}  
\end{figure*}

\begin{table*}
    \begin{tabular}{|c|c|P{\colwidth}|c|P{\colwidth}|P{\colwidth}|c|c|}

    \hline
    \textbf{Models} & \textbf{Tasks} & \textbf{Qubits} & \textbf{Channels} & \textbf{Depth} & \textbf{Gates} & \textbf{Parameters} & \textbf{MSE} \\
    \hline
    \multirow{3}{*}{QDE} & Cosine-wave &2 & 1 & 1 & 5 & 2 & $1.10\times 10^{-6}$ \\
    \cline{2-8}
    & Composite-periodic signal &4 & 2 & 1 & 10 & 4 & $5.21\times 10^{-6}$ \\
    \cline{2-8}
    & Composite-aperiodic signal &4 & 2 & 1 & 10 & 4 & $8.10\times 10^{-5}$ \\
    \hline
    QRC & All the above &3 & \NAline & \NAline & \NAline & 26 & See Fig.~\ref{fig: errorbar1} (\ref{fig: errorbar2} and \ref{fig: errorbar3}) \\
    \hline
    
    \end{tabular}
   
    \caption{\textbf{Quantum Resources Used in QDE and QRC Models.} 
    The QRC model is implemented through Hamiltonian evolution $e^{-iHt}$ rather than an explicit gate-based circuit. 
    Therefore, resource metrics used in the QDE model such as channel count, single-block depth, and gate count are not applicable to QRC and are indicated as ``\NAline". In the QDE model, comparisons are made by scaling the number of single $(1,1)$-QDE blocks, referred to as channels. 
    The ansatz employed is the HEA as described in Sec.~\ref{sec: settings}.}
    \label{tab:benchs1}
\end{table*}

\begin{figure*}[htbp]
    \centering
    \begin{tikzpicture}
        \node[anchor=center] (a) at (0,0) {\includegraphics[width=0.7\linewidth]{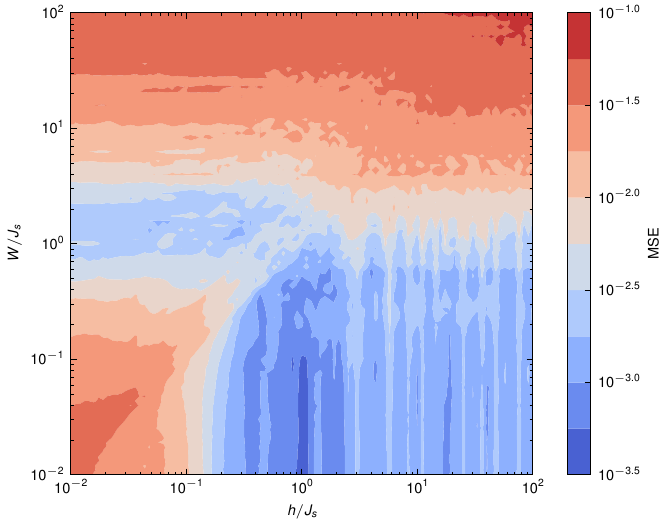}};
        
        \node[anchor=center] (b) at ($(a.south)-(0.25\linewidth,4cm)$) {\includegraphics[width=.48\linewidth]{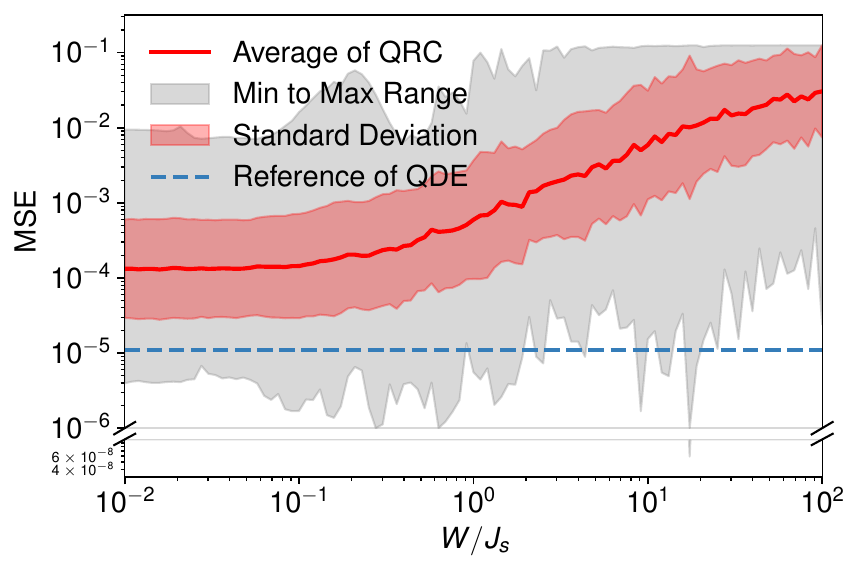}};
        
        \node[anchor=center] (c) at ($(a.south)+(0.25\linewidth,-4cm)$) {\includegraphics[width=.48\linewidth]{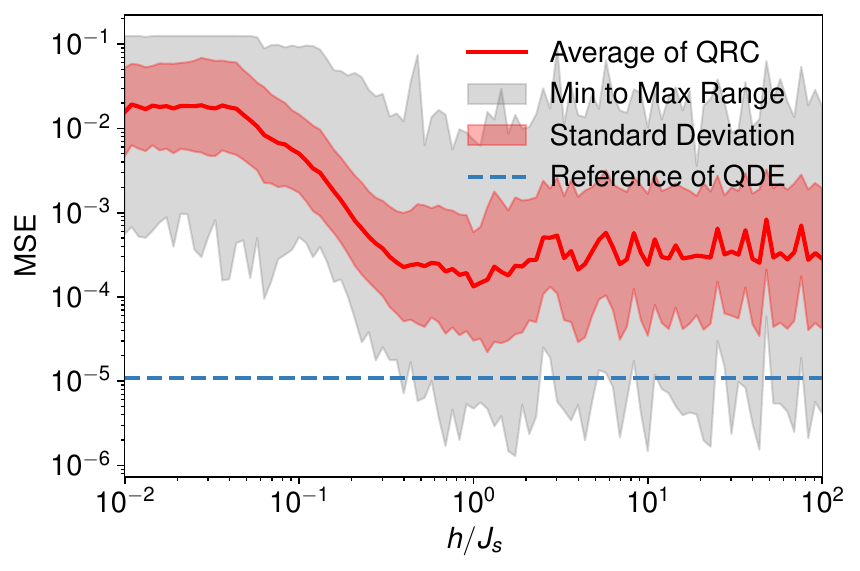}};
        
        \node[anchor=north west] at ($(a.north west)-(0.2cm,0.2cm)$) {\textbf{(a)}};
        \node[anchor=north west] at ($(b.north west)-(0.2cm,0.2cm)$) {\textbf{(b)}};
        \node[anchor=north west] at ($(c.north west)-(0.2cm,0.2cm)$) {\textbf{(c)}};
    \end{tikzpicture}
    \caption{\textbf{Comparison of QDE with QRC on Cosine-Wave Signal Prediction.} (a) Error map for different values of the Hamiltonian hyper-parameters of QRC, with the transverse field $h$ and the disorder bound $W$ varying logarithmically in the range $[10^{-2}, 10^2]$. Results are averaged over 100 realizations. (b) and (c) With $h/J_s$ ($W/J_s$) fixed at the minimum error from map (a), the other hyper-parameter $W/J_s$ ($h/J_s$) is varied. The average results are shown with the minimum to maximum error range (grey shadows) and standard deviation (red shadows). The reference line for the QDE model represents the average performance over different circuit parameter initializations using a single $(1,1)$-QDE channel. Since the QDE used here is based on a HEA architecture without hyper-parameters, it is naturally independent of the x-axis values.}
    \label{fig: errorbar1}
\end{figure*}

\begin{figure*}[!t]
    \centering
    \begin{tikzpicture}
        \node[anchor=north east] (img) at (0,0) {\includegraphics[width=1.0\textwidth]{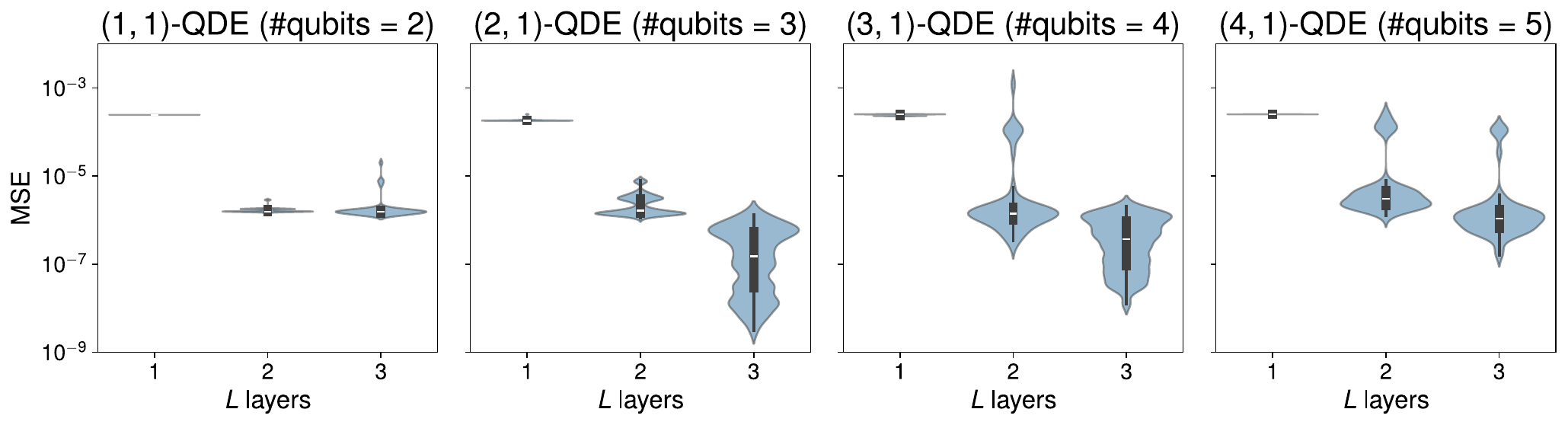}};
    \end{tikzpicture}
    \caption{\textbf{Performance of the QDE model on NARMA2.} The four subfigures are arranged from left to right by increasing qubit count. Each violin plot represents the distribution over 50 trials, consistent with previous tests.
    }
    \label{fig: NARMA2}  
\end{figure*}

This part establishes both numerical and experimental evidence on some applications to characterize the adaptability of the QDE introduced above. Firstly, we employ the QDE to predict a cosine-wave signal with a single channel. In the context of composite signal functions, our approach effectively captures the characteristics inherent in superpositions of different cosine-wave signals with the architecture depicted in Fig.~\ref{fig: illustr}(c).
To further demonstrate the QDE's versatility, we assess its denoising capability on signals exposed to uniform noise. 
For benchmarking purposes, we compare it with the performance using QRC. 
Due to the existence of local minima in the QDE optimization landscape, we report the median performance over multiple initial parameter sets. The distribution of these performances is illustrated using a violin plot, as shown in Fig.~\ref{fig: scaling}, taking the cosine-wave signal as an example. In contrast, the QRC model, trained via linear regression, demonstrates less sensitivity to initial parameters; thus, we report the average performance over 100 realizations varying only hyper-parameters, excluding training parameters.
Additionally, we focus on long-term time series and noisy simulations, training and forecasting them at a ratio of $1:10$ under the influence of depolarizing or amplitude-damping noise. Notably, an experimental test on a superconducting quantum computer indicates significant promise for real-world time series processing.

We employ the HEA architecture in \REVISE{these applications}~ and tailor the number of linear-composite channels to align with \REVISE{each } task's complexity. \REVISE{By default, to evaluate per-step predictive accuracy and enable fair comparisons, we report one-step predictions, i.e., at each time step the ground-truth value is provided as input, consistent with standard QRC evaluations. Nevertheless, because QDE is designed for multi-step forecasting, we also provide autoregressive (multi-step) simulation results and analyses in Secs.~\ref{sec: noisetype}--\ref{sec: analysis}. For datasets in which the input and target signals arise from different feature sets (e.g., NARMA), we report only one-step prediction results throughout this work.}

\paragraph{Cosine-wave and its composite} Firstly, we apply the QDE on the cosine-wave signal, $$x(t)=A\cos(\omega t+\phi),$$ where $\omega=0.04\pi$, $A=0.5$ and $\phi=0$ for simplicity. 
Then we consider the composite of Cosine-waves, $$x(t) = \sum_{i} A_i \cos(\omega_i t + \phi_i),$$ where each term represents an independent wave component. 
Two simplest non-trivial examples will be presented, a periodic and an aperiodic one with only two linear-composite channels. 
In the periodic case, we set $\omega_1 = 0.04\pi$ and $\omega_2 = 2\omega_1$; the aperiodic case, $\omega_1 = 0.04\pi$ and $\omega_2 = \sqrt{5}\omega_1$. 
We define the temporal domain as $0 \leq t \leq 200, t \in \mathbb{N}$. 
In this and the following demonstrations, we train the QDE using the first 100 data points generated from the equation (light red dots) and evaluate its extrapolation performance on the subsequent 100 points (blue triangles), as shown in Fig.~\ref{fig: applic}.
The ansatz used here is a simplified version of HEA, in which the three-parameter gate $U_1(\theta_1, \theta_2, \theta_3)$ is replaced with a single-parameter rotation $R_Y(\theta)$ to investigate the minimal quantum resources required for signal prediction.
We begin by identifying the lowest resource configuration for the three signals, as summarized in Table~I: one $(1,1)$-QDE channel is sufficient for the cosine wave, while two channels are used for the other two signals.
The prediction error, defined by the MSE and averaged on 50 samples with different initializations, is $1.10\times 10^{-6}$, $5.21\times 10^{-6}$ and $8.10\times 10^{-5}$ with the lowest resource, respectively, which demonstrates the capability of QDE to accurately represent the periodic nature of each component of the signal. 
Detailed theoretical foundations of these observations are further elaborated in Sec.~\ref{sec: analysis}.

Then, using the cosine-wave prediction as an example, we analyze the scalability of the QDE model by either increasing the total number of $(1,1)$-QDE channels or increasing the number of qubits within a single $(n_m, n_x)$-QDE block. The first method combines multiple independent $(1,1)$-QDE outputs classically without entanglement, while the second increases internal qubit count to enable entanglement within the block, creating a more complex quantum state. Due to local minima in optimization, violin plots are used to display the full MSE distribution. Results in Fig.~\ref{fig: scaling} show that increasing the channel number significantly reduces MSE, whereas the performance gains from increasing layers eventually saturate, dependent on the channel number. Similarly, increasing internal qubits improves best-case performance but with greater variability and instability.

\paragraph{Denoising of signals}
To evaluate the denoising capability of the QDE, we expose the aforementioned three signals to uniform noise randomly distributed within the range of $[-0.1, 0.1]$. 
Each noisy signal undergoes processing via the QDE to evaluate its efficacy in recovering the original, clean signal. 
The number of training points in these versions is increased to enhance the capture of signal characteristic mode buried in random noise.
For the cosine-wave, the QDE markedly reduces noise interference, achieving a denoising MSE of $3.07 \times 10^{-3}$, as illustrated in Fig.~\ref{fig: applic}(d). 
Likewise, the periodic and aperiodic composite signals demonstrate the QDE's robustness in noise reduction, recording MSEs of $3.97 \times 10^{-3}$ and $3.21 \times 10^{-3}$, respectively, as shown in Fig.~\ref{fig: applic}(e) and (f). 
The prediction part of each subplot is juxtaposed with theoretical clean signals for comparison. 
We define the deviation as $\delta  = \hat{x}(t)-x_{\text{ref}}(t)$, which represents the difference between the output of the QDE ($\hat{x}(t)$) and the reference signal value ($\hat{x}(t)$).
Although $\delta$ is random and large during the training stage, as indicated by the red dashed line, the QDE effectively extracts the noise-free mode from the signal, showcasing its notable performance in denoising.
These results not only underscore the QDE's effectiveness in signal fidelity restoration but also highlight its utility in practical noise reduction scenarios.

\paragraph{Comparison with QRC}
In this application, we compare the performance of the QDE and QRC models on the aforementioned tasks. 
For the QRC model, we utilize a 3-qubit system with 5 evolution slices ($V=5$), keeping the training and prediction lengths consistent with those of the QDE model. 
The results of this comparison for the first task, cosine-wave signal prediction, are shown in Fig.~\ref{fig: errorbar1}(a)-(c). 
The error map for different values of the Hamiltonian hyper-parameters of QRC is plotted with the transverse field $h$ and the disorder bound $W$ varying logarithmically in the range $[10^{-2}, 10^2]$.
Results are averaged over 100 realizations. 
In Fig.~\ref{fig: errorbar1}(b)-(c), we fix $h/J_s$ (or $W/J_s$) at the optimal value identified in Fig.~\ref{fig: errorbar1}(a). The other parameter ($W/J_s$ or $h/J_s$) is varied within the range $[10^{-2}, 10^2]$. 
The median MSE of the QDE model is calculated from 50 random initializations, as detailed in Fig.~\ref{fig: scaling} and is indicated by the dashed light blue line. The median MSE consistently remains below the QRC model's average error (red curve) and standard deviation (red shaded area).
Detailed configuration parameters and resource calculations for both models are summarized in Table~\ref{tab:benchs1}, representing the minimal resources required to accomplish the task.
Despite some instances of lower prediction error across 100 realizations, the 2-qubit QDE model exhibits more stable performance in terms of precision, circuit depth, and training resource allocation compared to the QRC model. 
Further results on the other two tasks, along with detailed insights into the QRC theory and its configurations, can be found in the supplementary information.

\paragraph{NARMA2} To further demonstrate the capability of the QDE model on more complex time series, we apply it to the nonlinear autoregressive moving average (NARMA) task. A typical NARMA signal of order $n$ is generated by the following dynamics:
$$x_{t+1}=...\alpha y_t+\beta y_t\left(\sum_{j=0}^{n-1}y_{t-j}\right) +\gamma s_{t-n+1}s_t+\delta, $$
where $x_t$ denotes $x(t)$ for simplicity, and the parameters $(\alpha, \beta, \gamma, \delta)$ are set to $(0.3, 0.05, 1.5, 0.1)$. The input signal $s_t$ is superimposed sine wave, following the setup in \cite{Fujii2017}:
$$s_t = 0.1\left[ \sin\left(\frac{2\pi a t}{T}\right) \sin\left(\frac{2\pi b t}{T}\right) \sin\left(\frac{2\pi c t}{T}\right)  + 1\right], $$
with $(a,b,c)=(2.11, 3.73, 4.11)$ and $T=100$. In this simulation, we slightly modify the input and output protocol, as shown in Fig.~\ref{fig: illustr}, to process the input $(\bm{m}_t, s_t)$ to output $(\bm{m}_{t+1}, x_{t+1})$, aligning with the settings in \cite{Fujii2017}. We set $n=2$, train for 500 steps, and predict for 100 steps.
Given the complexity of this time series, we chose to scale the model by increasing the number of internal qubits, rather than simply adding more channels, and by deepening the ansatz layers to fully exploit entanglement among qubits, aiming to improve model expressiveness. All results are over 50 trails with different initializations. Results using the ansatz HEA for NARMA2 are shown in Fig.~\ref{fig: NARMA2}, with a demonstration of a $(2,1)$-QDE provided in Fig.~\ref{fig: applic}(g).
We observe that increasing the number of ansatz layers consistently reduces the error, while increasing qubit count does not always guarantee better performance due to increased optimization complexity. This example highlights the QDE model's potential for handling more realistic and challenging time series beyond simpler composite signals. For other complex tasks like learning and predicting the nonlinear dynamics of the Rayleigh equation, please refer to the supplementary information.

\begin{figure*}[htbp]
    \centering
    \begin{tikzpicture}
        \node[anchor=center] (a) at (-1.5cm,0) {\includegraphics[width=0.5\linewidth]{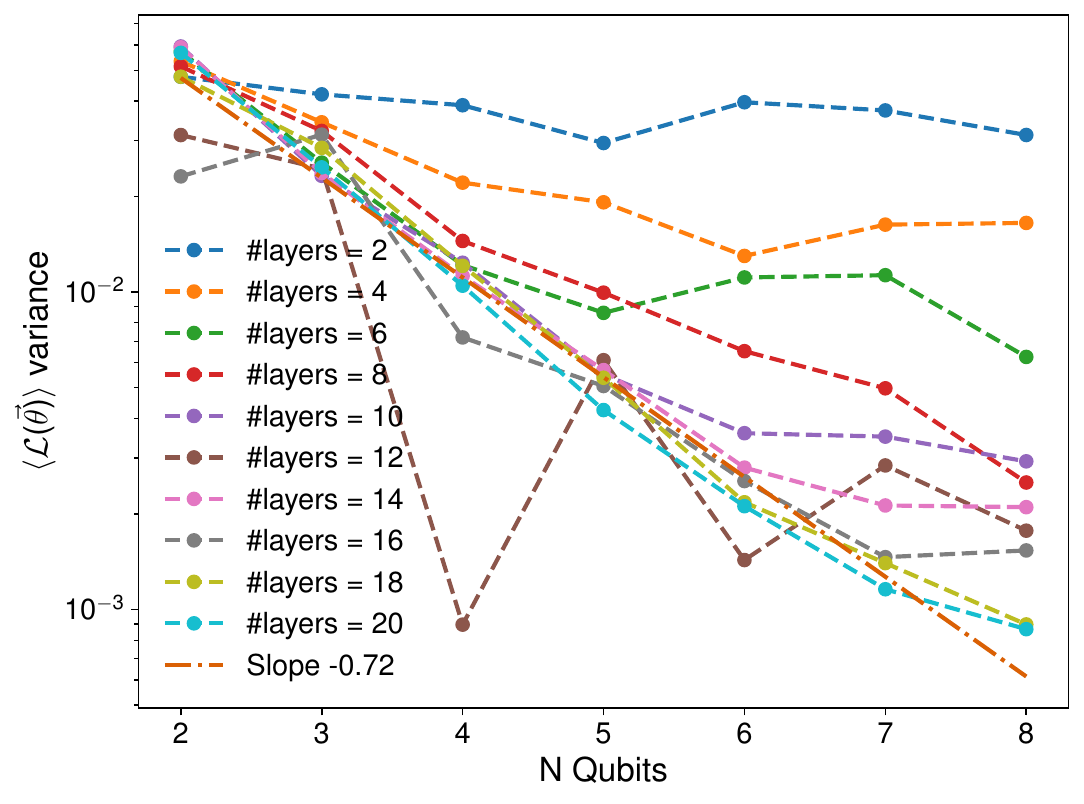}};
        
        \node[anchor=center] (b) at (7.5cm, 0) {\includegraphics[width=.5\linewidth]{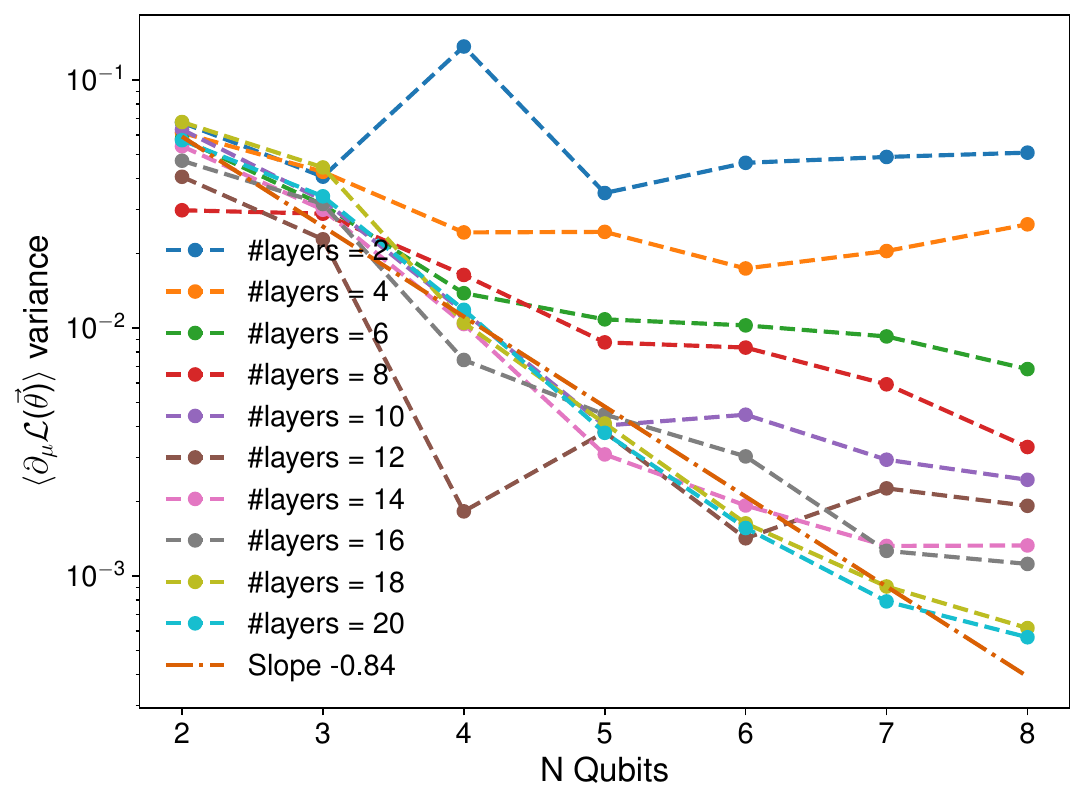}};
        
        \node[anchor=north west] at ($(a.north west)+(0.4cm, -0.2cm)$) {\textbf{(a)}};
        \node[anchor=north west] at ($(b.north west)+(0.4cm, -0.2cm)$) {\textbf{(b)}};
    \end{tikzpicture}
    \caption{\textbf{Barren Plateau Evaluation of the $(n_m,1)$-QDE Model Using the HEA} The total qubit number is scaled from 2 to 8. The variance of the loss function $\langle \mathcal{L}(\bm{\theta})\rangle$ (a) and its gradient $\langle \partial_{\mu} \mathcal{L}(\bm{\theta})\rangle$ for specified index $\mu$ (b), are shown varying with qubit number for different ansatz layer depths and plotted with a semi-log style. Each data point represents an average over 400 samples of parameters drawn uniformly from $[0,2\pi)$.}
    \label{fig: BP}
\end{figure*}

\paragraph{The Problem of Barren Plateau} As discussed in Sec.~\ref{sec: gradient}, the potential issue of barren plateaus in the QDE model is an important consideration. To investigate this, we use the NARMA2 signal as a test case and compute both the variance of the loss function and the variance of its gradients. The circuit setups are consistent with those described previously. The results, shown in Fig.~\ref{fig: BP}, present these variances averaged over 400 samples with parameters $\theta_i\in[0, 2\pi)$.
From Fig.~\ref{fig: BP}, we observe that when the number of ansatz layers is small, the exponential decay of variance with increasing qubit number is not apparent. However, as the number of layers increases, this decay trend becomes more evident, and at 20 layers, the data closely follow an exponential decay pattern consistent with barren plateau behavior. This indicates that the QDE model is susceptible to barren plateaus when the ansatz depth is large.
Although the layer depths used in our current experiments are not sufficient to fully enter the barren plateau regime, the phenomenon remains a limitation of the model. Addressing methods to mitigate or avoid barren plateaus is beyond the scope of this study and will be the focus of our future research efforts.

\begin{figure*}[!th]
    \centering
    \includegraphics[width=1.0\textwidth]{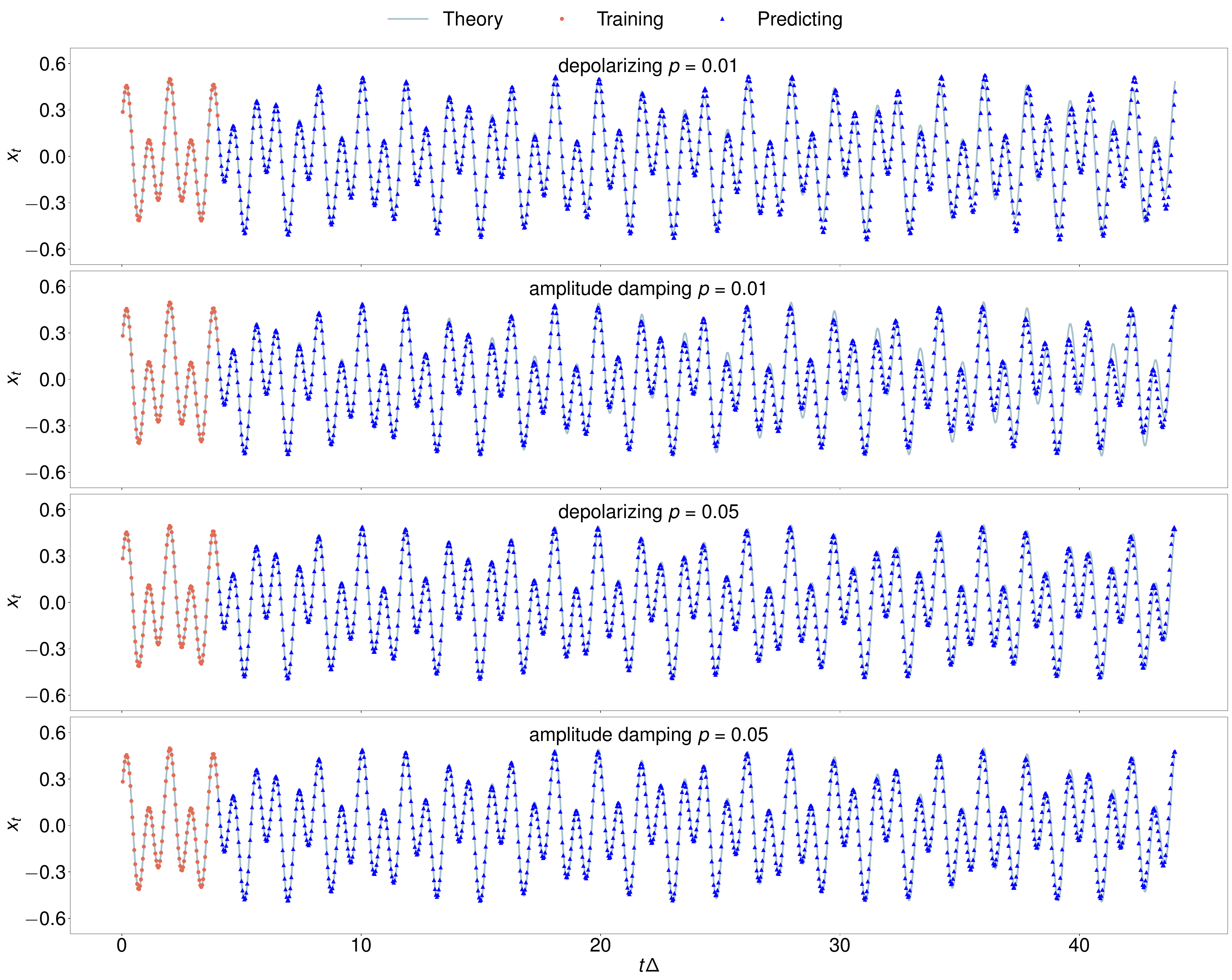}
    \caption{\textbf{Long-term Prediction on Composite Aperiodic Signal with Quantum Noise.} The training and predicting protocols replicate those in Fig.~\ref{fig: applic}, with the distinction of a training-to-predicting ratio of $1:10$.  
    }
    \label{fig: longterm3}
\end{figure*}

\subsection{Performing long-term prediction by recovering from noisy results} \label{sec: noisetype}

In this part, we show how the QDE can be applied to noisy quantum computers. 
As analyzed above, the long-term coherence of qubits need to be kept for the QRC and QRNN models. 
As a result, the long-term prediction task turns to be formidable due to the strong sensitivity to noise. Here, we propose a trainable method to recover the system from noisy results. Moreover, we show that our method can allow the QDE to spontaneously unbias the difference among different quantum computers.

First, we analyze the effect of noise for a QDE running on a noisy quantum computer with typical error sources, including the depolarizing, the amplitude-damping, and the readout error. We denote the noiseless and actual results for qubit $i$ as $\langle \sigma_z^i\rangle^*$ and $\langle \sigma_z^i\rangle$, respectively. 

For the depolarizing noise, let $\sigma$ be the initial quantum state, $\rho^{(k)}$ the state after $k$ applications of the depolarizing and unitary channels, and $\rho$ the state absent noise channels. We can express $\rho^{(k)}$ as: 
\begin{equation}
\rho^{(k)} \equiv\left[\left(\prod_{j=1}^{k} \mathcal{U}_{j} \circ \mathcal{E}^{\mathcal{D}}_{p_{j}}\right)\right] \circ \sigma,
\end{equation}
where $p_j$ represents the probability of the depolarizing channel $\mathcal{E}^{\mathcal{D}}_{p_{j}}$ and $\mathcal{U}_j(\cdot)=U_j \cdot U_j^\dagger$ represents the unitary channel. 

Drawing from Lem.~1 in~\cite{LaRose2020}, this is reformulated as:
\begin{equation}
\begin{aligned}
\rho^{(k)}=&\left( \prod_{j=1}^{k} (1-p_{j}) \right) U_{k}\cdots U_{1} \sigma U_{1}^{\dagger} \cdots U_{k}^{\dagger}\\
&+\left(1-\prod_{j=1}^{k} (1-p_{j})\right) \frac{I}{2^n}\\
=&\prod_{j=1}^{k} (1-p_{j}) \rho +\left(1-\prod_{j=1}^{k} (1-p_{j})\right)\frac{I}{2^n}.
\end{aligned}
\end{equation}

Subsequently, the noisy result $\langle \sigma_z^i\rangle$ is represented as:
\begin{equation}
\begin{aligned}
\langle \sigma_z^i\rangle= &\operatorname{Tr}[\rho^{(k)}\sigma_z^i]\\
=&\left( \prod_{j=1}^{k} (1-p_{j}) \right) \langle \sigma_z^i\rangle^*.
\end{aligned}
\end{equation}

For the amplitude damping noise, characterized by the amplitude damping channel $\mathcal{E}_{\gamma}^{\mathcal{AD}}$, it affects the quantum state $\rho$ like:
\begin{equation}\label{eq:dep}
\begin{aligned}
\langle \sigma_z^i\rangle= &\operatorname{Tr}[\mathcal{E}_{\gamma}^{\mathcal{AD}}(\rho)\sigma_z^i]\\
=&(1-\gamma)\langle \sigma_z^i\rangle^*+\gamma,
\end{aligned}
\end{equation}
where $\gamma$ is the strength of the amplitude damping channel. 

For the readout error, this can be understand as a probabilistic transformation of ideal measurement results. Specifically, the readout error can be characterized as bit-flip events, which is quantified using a response matrix $M$. The observed outcome probabilities, denoted as $\bm{p}_{\operatorname{obs}}$, are obtained from the transformation of the error-free probabilities $\bm{p}_{\operatorname{ideal}}$ as follows:
\begin{equation}
\bm{p}_{\operatorname{obs}}=M\bm{p}_{\operatorname{ideal}}.
\end{equation}
Given that the observable $\sigma_z^i$ can be decomposed into projectors of computational bases, it is feasible to express the result with readout error as the combination of elements from $\bm{p}_{\operatorname{obs}}$, further as the linear transformation on the ideal result. This is similar to what is observed in depolarizing and amplitude damping scenarios.

To enhance the recovery of results from noisy quantum computers, we introduce a linear layer called LECL, consisting of a linear transformation with parameters $A:\mathbb{R}^{n\times n}$, $\mathbf{b}:\mathbb{R}^{n}$. 
This layer performs the mapping $(\vecx, \vecm)^T\mapsto A(\vecx, \vecm)^T + \mathbf{b}$ on the obtained measurements within $n^2 + n$ parameters. The computational complexity of training and inference with this linear layer remains within $\mathcal{O}(n^2)$, facilitated by the implementation of the back-propagation algorithm~\cite{Rumelhart1986}. Notably, since this procedure operates within classical computational frameworks, it does not increase the number of executions required for the quantum circuits.
We emphasize that the quantum process should still be the resource of the ability of this model, instead of this linear layer. A model comprising only the linear layer, without the QDE, lacks non-linearity and is therefore unsuitable for most problems.

We evaluated long-term prediction capabilities of the QDE under aforementioned noise channels using composite Cosine-wave signals. 
\REVISE{To demonstrate long-term forecasting performance and robustness to accumulated errors, here we employ a multi-step (autoregressive) prediction protocol for composite signals: each newly predicted value is fed back as input for the next time step. }
The aperiodic signal is shown in Fig.~\ref{fig: longterm3}, with additional examples provided in the supplementary material. 
Maintaining a training-to-predicting ratio of $1:10$ and replicating previous settings, including the LECL, these simulations highlight the robustness of the QDE complemented by a linear layer. This robustness demonstrates the QDE's potential for long-term time series prediction in practical scenarios.

The numerical results also demonstrate the models' stability under different noise strengths, showcasing the main benefit of this error-recovery QDE method: the system is trained not only to capture dynamical features but also to adapt to the system's noise characteristics. Due to the efficiency of the training algorithm, we achieve enhanced adaptability without incurring additional costs.

\begin{figure*}[htbp]
    \centering
    \begin{tikzpicture}
        \node[anchor=center] (center) at (0,0) {};
        \node[anchor=east] (l) at ($(center.east)+(0,0)$) {\includegraphics[width=0.45\linewidth]{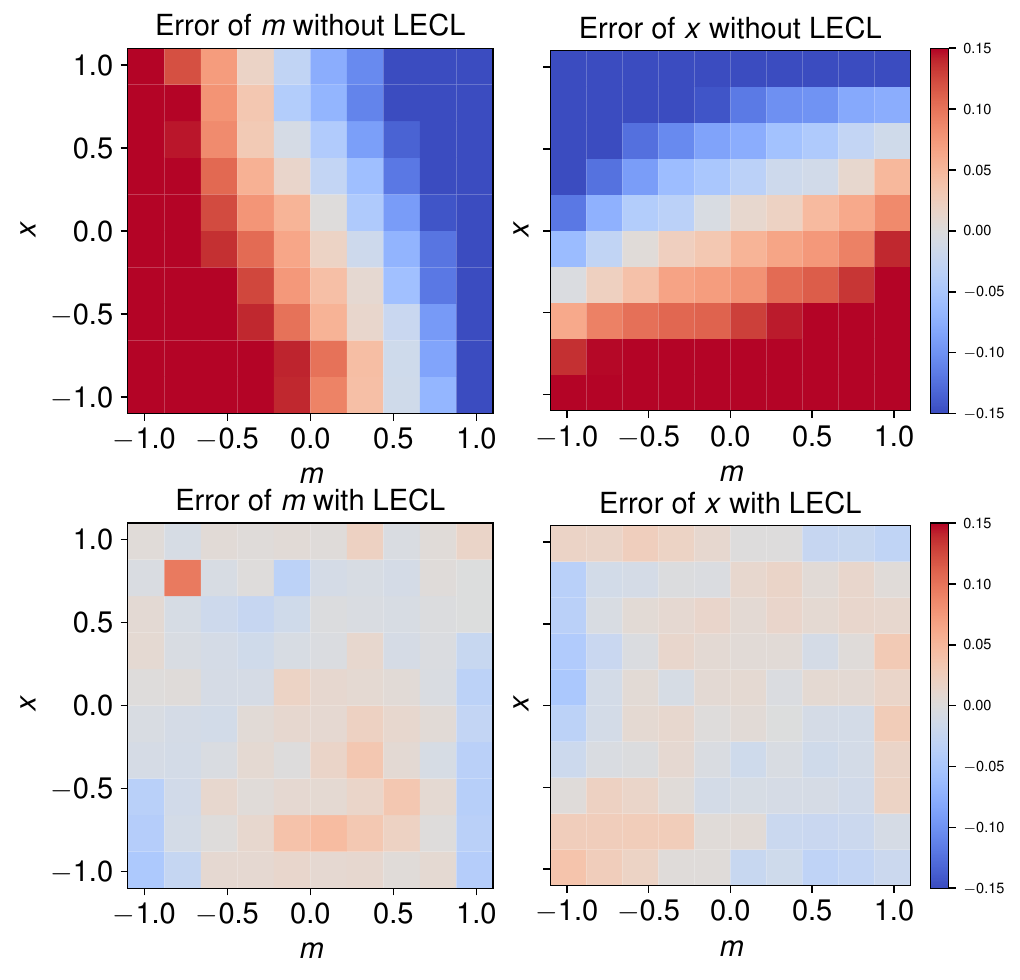}};
        
        \node[anchor=west] (r) at ($(center.east)+(0,0)$) {\includegraphics[width=.5\linewidth]{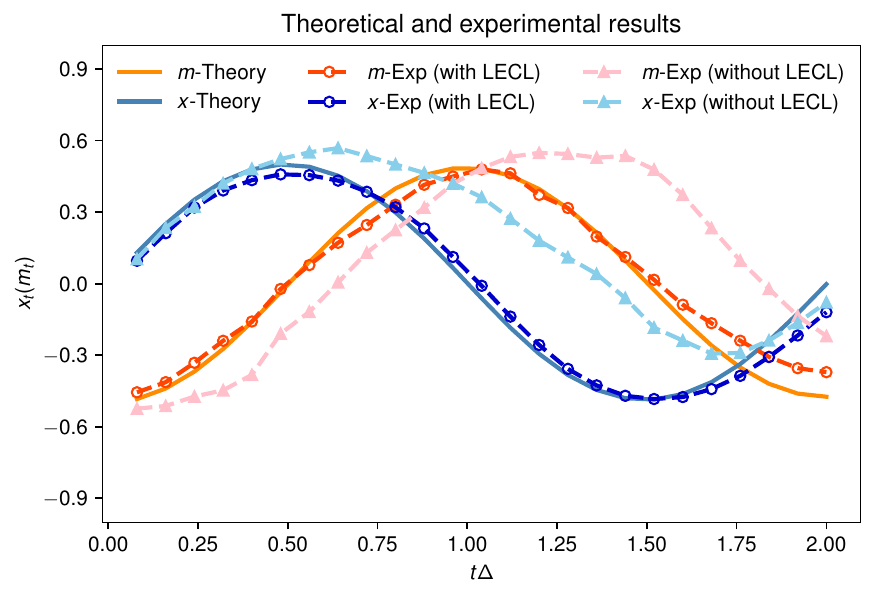}};
        
        \node[anchor=north west] at ($(l.north west)-(0.2cm,0.23cm)$) {\textbf{(a)}};
        \node[anchor=north west] at ($(l.north west)-(0.2cm, 4.cm)$) {\textbf{(b)}};
        \node[anchor=north west] at ($(r.north west)-(0.1cm,0.2cm)$) {\textbf{(c)}};
    \end{tikzpicture}
    \caption{\textbf{Error Distribution in Encoding Space and the Cosine-wave Signal Generation.} 
    (a) Error distribution within the encoding space $(m,x)$, each grid point indicating the deviation between the ideal output $(m^*, x^*)$ and the quantum computer's output $(\hat{m}, \hat{x})$ without LECL, denoted as $\epsilon_m=\hat{m}-m^*$ and $\epsilon_x=\hat{x}-x^*$. 
    (b) Adjusted error distribution following the application of LECL, illustrating the changes in error magnitude. 
    (c) Generation of a single period of a cosine-wave time series, showing the data register dynamics and the memory register responses; solid lines correspond to the results from ideal simulations with optimized parameters, while dashed lines represent the outcomes from the quantum computer after the LECL is applied.}
    \label{fig: exp}
\end{figure*}

\subsection{Experimental demonstration on a superconducting quantum processor} \label{sec: exp}

As previously discussed, the realization of time series predictions is challenged by different typical errors. Our approach addresses these compounded errors by focusing on the mitigation of gate and readout errors with LECL method. This focus is crucial for maintaining accuracy in our model, as the fixed circuit depth does not increase with the time series length, making gate and readout error mitigation paramount regardless of decoherence.

Previous studies have developed error mitigation techniques to refine expectation values on quantum computers~\cite{Temme2017,Endo2018,Kandala2019, Alistair2021, Strikis2021}. 
In this experimental demonstration on the Origin ``Wukong'' platform, we utilize a 2-qubit system for cosine-wave prediction. 
We leverage a learning-based error mitigation approach, integrated with learnable error-cancellation layer as previously described, considering only gate and readout errors. 
Different from the direct error model determination and probabilistic correction in~\cite{Temme2017,Endo2018}, we aim to average errors within the encoding space, establishing a transformation between noiseless results and those obtained from near-term quantum computers. 
This transformation is vital in error compensation during time series forecasting.

To further elaborate, we consider executing quantum circuits with grid inputs $m^i,x^i$ in the state space $m, x \in [-1,1]$. The discrepancy between the ideal outputs, $(m^*, x^*)=\QDE(m,x)$, and actual outputs from the quantum processor without LECL, $(\hat{m}, \hat{x})$, is quantified as $\epsilon_m=\hat{m}-m^*$ and $\epsilon_x=\hat{x}-x^*$. Subsequently, an LECL characterized by parameters $A$ and a bias vector $\mathbf{b}$ is applied to the experimental results within this state space. These parameters are optimized with Broyden-Fletcher-Goldfarb-Shanno (BFGS) algorithm~\cite{Broyden1970,Fletcher1970,Goldfarb1970,Shanno1970}, implemented in SCIPY~\cite{SciPy2020}, and fixed for the time series generation. Utilizing these optimized parameters, we construct a $25$-step time sequence \REVISE{generated autoregressively to span a single period of the cosine wave: only the first input is provided, and the time sequence is uniquely determined by the first input}.

The experiments are performed on the ``Wukong'' quantum processor of the Origin platform and reveal the error distribution with and without LECL, depicted in Fig.~\ref{fig: exp}(a) and (b). Comparative analysis of error distributions before and after applying the optimization procedure reveals that errors are reduced to below the threshold of $10^{-2}$. Consequently, the time series can be reconstructed using the optimized linear transformation. As demonstrated in Fig.~\ref{fig: exp}(c), this error mitigation technique effectively prevents divergence in the time series data, validating the feasibility of the QDE on NISQ platforms and marking a significant advance in quantum-based time series analysis.

\subsection{Dynamics analysis}\label{sec: analysis}
\begin{figure}[!tb]
    \label{fig: phaseplot}
    \centering
    \includegraphics[width=0.95\linewidth]{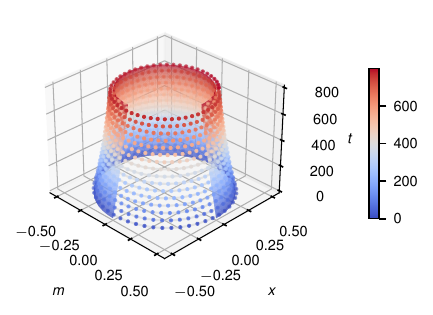}
    \caption{\textbf{Comparison of QDE Trajectory and Its Linear Approximation.} The trajectory generated by the QDE of Eq.~\eqref{eq:QDE1} extends along the timeline, while the surface is plotted based on the dynamics of Eq.~\eqref{eq:rotation1} with parameters set to $(a^*,b^*,\gamma^*)$.}
    \label{fig: linearapprox} 
\end{figure}

The classical dynamics exhibit a variety of characteristics, whereas the properties of quantum dynamical embedding remain less explicit. An in-depth analysis of classical dynamical properties, such as periodicity and fixed points, facilitates a foundational understanding of the QDE and enhances the learnability of data-driven methods. Nevertheless, dynamical analysis necessitates an explicit formulation of the QDE, a task that becomes increasingly complex with higher qubit number or greater circuit depth. 

Utilizing a simplified $(1,1)$-QDE architecture with HEA construction which is used in Sec.~\ref{sec: applic} for cosine-wave prediction, we derive the theoretical mapping equations as follows:

\begin{equation}\label{eq:QDE1}
    \left\{ \begin{array}{ll}
    m_{t+1}&=m_t \cos \theta_1 -x_t\sqrt{1-m^2_t}\sin \theta_1,\\
    x_{t+1}&= x_t\cos \theta_2- m_t\sqrt{1-x^2_t}\sin \theta_2.
    \end{array}  \right.
\end{equation}

In subsequent analysis, we treat the given data $\vecx_{t+1}$ and its estimator $\hat{\vecx}_{t+1}$ as equivalent within a generative model framework, not distinguishing between them. Although quantum computation operates linearly with unitary quantum operators, it is possible to derive a nonlinear discrete map of the state vector. This nonlinearity, evident as a square root in Eq.~(\ref{eq:QDE1}), arises from encoding and measurement protocols. Nonetheless, analyzing nonlinear dynamical systems typically involves identifying fixed points and linearizing the system in a small neighborhood, a technique that loses efficacy when initial values are distant from these points, limiting its applicability across the entire phase plane. While identifying the fixed point at $(0,0)$ of $(1,1)$-QDE is straightforward, linear properties in its neighborhood can not extend to the global phase space.

Considering a specific scenario where $\theta_2=-\theta_1=\theta>0$ and $x,m\in [-0.5,0.5]$, the square root terms in Eq.~\eqref{eq:QDE1} range from $0.866$ to $1.0$. To linearize the equations, we replace the square root with a constant $\gamma \in (0.866, 1.0)$, transforming Eq.~(\ref{eq:QDE1}) into a matrix form $(m_{t+1},x_{t+1})^T =R(m_t,x_t)^T$, where $R$ is the Jacobi matrix:
\begin{equation}\label{eq:rotation1}
R=b\begin{bmatrix}
\cos a&\gamma \sin a\\
-\gamma \sin a & \cos a
\end{bmatrix},
\end{equation}
with $a$ being an angle correction and $b$ an amplitude correction relative to $1$. The eigenvalues of $R$ are $\lambda_\pm =b(\cos a\pm i\gamma\sin a)$. Optimizing $(a,b,\gamma)$ for an initial point $(m_0,x_0) =(0,0.5)$ and a fixed $\theta=0.04\pi$ (corresponding to the angular frequency of a cosine-wave, as discussed in Sec.~\ref{sec: applic}), we find that $|\lambda_\pm| <1$ for optimal values $(a^*,b^*,\gamma^*)=(0.12754,0.9996,0.9973)$. This implies that the fixed point $(0,0)$ is a stable focus, and the system's dynamics will converge to it over extended evolution.

A more concise expression of the discrete map can be achieved by rescaling the matrix as $R=\beta\tilde{R}$, where $\beta =b\sqrt{\cos^2a+\gamma^2\sin^2a}<1$ and $\tilde{R}$ is an orthogonal matrix:

\begin{equation}\label{eq:rotation2}
\tilde{R}=\begin{bmatrix}
\cos \alpha& \sin \alpha \\
-\sin \alpha & \cos \alpha
\end{bmatrix},
\end{equation}
with $\alpha=\arctan(\gamma\arctan(a))$. For a given initial point $(m_0,x_0)^T$, the state after $t$ iterations is $(m_t,x_t)^T = \beta^t \tilde{R}^t (m_0,x_0)^T$, leading to a norm $|(m_t,x_t)^T| =|\beta^t \tilde{R}^t (m_0,x_0)^T|=\beta^t |(m_0,x_0)^T|$. This reveals that the trajectory on the phase plane follows a logarithmic spiral, as illustrated in Fig.~\ref{fig: linearapprox}. 
The trajectory generated by the QDE is bounded by the surface formed by the linear dynamics of Eq.~(\ref{eq:rotation1}), validating the reasonability of our linear approximation in the presence of weak nonlinearity.

Adjusting the observables of quantum circuits allows for the simulation of similar dynamics. Introducing a scalar $\mu$ to the right-hand side of Eq.~(\ref{eq:QDE1}), the Jacobi matrix becomes $\mu R=\mu \beta\tilde{R}$, yielding eigenvalues $\lambda_\pm =\mu\beta(\cos \alpha\pm i\gamma\sin \alpha)$. By tuning $\mu$, we can ensure the eigenvalue norms are equal to or greater than $1$, indicating that the fixed point is either a center or an unstable focus, respectively. 

The preceding analysis elucidates the theoretical basis for the $(1,1)$-QDE model's ability to simulate the cosine-wave. This capability primarily stems from the QDE's weak nonlinearity, which, when combined with non-Markovian memory, enables a precise approximation of the cosine-wave signal.

\subsection{Universality}\label{sec: universal}
In this section, we explore the applicability of the QDE framework in approximating composite cosine-wave signals, which serve as a foundational step toward universality. The extension of QDE to model more complex dynamical systems remains an open question and a subject for future research.
We begin by examining QDE’s capability to model cosine-wave signals, governed by the equations in Eq.~(\ref{eq:QDE1}).

Here, $\theta_1$ and $\theta_2$ are parameters that can be adjusted to model the signals $x(t)=A\cos (\omega t)=A\cos (\pi \Delta t)$ and $m(t)=A\sin (\omega t)=A\sin (\pi \Delta t)$, subject to amplitude $A<1/2$, $t\in\mathbb{N}$ and time step length $\Delta<1/20$. Let $\theta_1 = -\theta<0$ and $\theta_2 = \theta>0$, then we have the following lemmas.

\begin{lemma}\label{lem:1}
    From a precise step $m(t)= \hat{m}(t)$ and $x(t)= \hat{x}(t)$, the error produced in the next step is less than $|\pi\Delta|/4$.
\end{lemma}

\begin{lemma}\label{lem:2}
    For an erroneous step $\hat{m}(t) = m(t) + \delta_m$ and $\hat{x}(t) = x(t)  + \delta_x$, suppose $|\pi \Delta| < 1/4$, the error produce in the next step is less than $\delta+\frac{1}{4}|\pi \Delta|$ where $\delta = \max(|\delta_m|, |\delta_x|)$. 
\end{lemma}

Proofs of Lem.~\ref{lem:1} and Lem.~\ref{lem:2} can be found in the supplementary information (See Supplementary Detailed Proofs).

These two lemmas establish the error propagation from both an error-free initial point and an erroneous initial point to the next time step. Building on these foundational lemmas, we present the following theorems:

\begin{theorem}[Single mode cosine-wave approximation]\label{thm:1}
    By properly selecting $\theta_1$ and $\theta_2$, QDE can approximate $A\cos (\pi\Delta t+\phi)$, where $|A|< 1/2$, $|\pi\Delta|<1/4$ with a bounded error increasing linearly over time steps.
\end{theorem}

\begin{theorem}[Composite cosine-wave approximation]\label{thm:2}
    Given a composite signal formed by the superposition of multiple cosine-waves, $\sum_i^K A_i\cos(\omega_i t+\phi_i) $, where each term has a distinct amplitude $A_i$, frequency $\omega_i$ and phase $\phi_i$. A $K$-channel QDE can approximate this composite signal with a bounded error by properly selecting parameters $\theta_1^1,\theta_2^2,\dots , \theta_1^K,\theta_2^K$, with each superscript indexing the corresponding channel. 
\end{theorem}

In Thm.~\ref{thm:1}, we proved the QDE's capability to approximate a single mode cosine-wave signal with a bounded error. This establishes the potential for QDE to approximate composite cosine-wave signals, as further explored in Thm.~\ref{thm:2}. Proofs are provided in the supplementary information (supplementary Detailed Proofs).

Lastly, we present a theorem that shows the universality of the QDE based on preceding discussions:

\begin{theorem}[Approximating any continuous function]\label{thm:3}
    If $g(t)$ is continuous on an interval $[a,b]$, then there exists a $K$-channel QDE approximating $g(t)$ with a bounded error for any $t$ in $[a,b]$. 
\end{theorem}

The proof is located in the supplementary information.

Through these lemmas and theorems, we underscore the universality of QDE in accurately modeling a wide range of signals, thereby reinforcing its potential in time series prediction.

\section{Discussion}\label{sec: disc}
In this research, we have developed a data-driven method that unites quantum machine learning with the analysis of dynamical systems in discrete time. The QDE provides a perspective of dynamical maps for understanding quantum systems and the characteristics of time series.

Distinct from existing quantum-based methods for time series prediction, our method offers a practical solution for near-term quantum computers. 
It addresses the significant challenge of quantum decoherence, a common obstacle in deep quantum circuits, by employing the QDE within a fixed circuit depth. 
Notably, the ansatz depth $C$ is significantly smaller than that required by full-sequence encoding methods, such as QRC, which scale as $C \times L$. Since the ansatz can be tailored to specific hardware constraints, QDE is particularly well-suited for current NISQ devices. 
This work represents an exploration of new effective strategies for temporal series processing, along with an acknowledgment of their limitations.

Through numerical simulations and practical experiments, our method demonstrates both long-term prediction capability and noise resilience in realistic scenarios. 
However, our simulations also reveal inherent optimization challenges, including the presence of local minima, the increased instability as the quantum system scales, and the barren plateau when the circuit depth of a single QDE block is sufficiently large.
Additionally, the current methods for data extraction and injection within QDE do not preserve the full entanglement structure through each time step. 
This leads to inevitable information loss compared to approaches like the QRC model, which employs a quantum reservoir to maintain memory. 
The numerical results, while highlighting the limitations of this approach, raise important questions about whether both this method and previous ones offer quantum advantages in the domain of temporal processing, and how their classical simulability is affected given the reduced quantum resources required for the corresponding tasks. Addressing these questions is beyond the scope of the present work and will be the focus of our future researches.

Looking forward, the QDE framework provides a promising foundation for future investigations into leveraging quantum circuits with smaller depth to capture complex system behaviors. While our results demonstrate encouraging progress in integrating quantum time series algorithms with dynamical system analysis, further research is necessary to overcome current limitations and fully understand the physical characteristics inherent in long-term time series prediction.

\section{Acknowledgements} 
We thank Song-Xin Qu for his valuable assistance with coding techniques and technical support.
This work was supported by National Key Research and Development Program of China (Grant No. 2023YFB4502500) and National Natural Science Foundation of China (Grant No. 12034018). 

\section{Data availability}
The data that support the findings of this study are available upon reasonable request from the authors.

%

\clearpage
\setcounter{table}{0}
\renewcommand{\thetable}{S\arabic{table}}%
\setcounter{figure}{0}
\renewcommand{\thefigure}{S\arabic{figure}}%
\setcounter{section}{0}
\setcounter{equation}{0}
\renewcommand{\theequation}{S\arabic{equation}}%

\onecolumngrid

\begin{center}
{\large \bf Supplementary Information\\
Data-driven Quantum Dynamical Embedding Method for Long-term Prediction on Near-term Quantum Computers}\\
\vspace{0.3cm}
\end{center}

\setcounter{page}{1}

\section{Quantum Reservoir Computing} \label{sec: qrc}
Quantum Reservoir Computing (QRC) is a useful approach for processing time sereis data on quantum computers. In this framework, each node is defined as a quantum orthogonal basis in Hilbert space. For a quantum state described by a density matrix in a system with $N $ qubits, there are $4^N$ elements. This quantum state equivalently represented as a classical state vector $\vec{x}$ with a dimensionality of $4^N$, which is similar to classical reservoir computing. 
The state injection in QRC is described as a completely positive and trace preserving (CPTP) map on the system, represented by the transformation $\rho_{t-1}\rightarrow \rho_s\otimes \mathrm{Tr}_s[\rho_{t-1}]$, where $\rho_s$ is the subsystem used for state encoding. Then the quantum reservoir envolves under a Hamiltonian $H$, which is expressed as:
\begin{equation}\label{eq: Hamiltonian}
    H = \sum_{i=1}^n (h+D_i) \sigma_{x,i} + \sum_{i=2}^n \sum_{j=1}^{i-1} J_{ij}\sigma_{z,i}\sigma_{z,j},
\end{equation}
where the disorder strength $D_i$ and the coupling strength $J_{ij}$ are randomly selected from uniform distributions in the interval $[-W, W]$ and the interval $[-J_s, J_s]$, respectively. The transverse strength $h$ and the disorder bound $W$ will be expressed in units of coupling bound $J_s$ and $J_s=1$ for convenience. All coefficients remain fixed during training as shown in~\cite{Fujii2017, Martinez2021}. Time multiplexing is employed to enhance learning, dividing the time interval $\tau$ into $V$ equal subintervals to increase the number of virtual nodes. The quantum state after $v$ subintervals' evolution is given by:
\begin{equation}
    \rho\left((t-1+v/V)\tau \right)=e^{-iH\tau v/V}\rho_{s_t}\otimes \mathrm{Tr}\left[\rho\left((t-1)\tau\right)\right]e^{iH\tau v/V}, v\in \{1,2,\dots , V\}.
\end{equation}
Computational nodes are obtained by measuring the quantum reservoir at the end of each subinterval using a selected measurement ensemble $O_z:=\{\sigma_{z,i},\sigma_{z,i}\otimes \sigma_{z,i+1}\}$. 
The QRC output is then obtained from a linear combination of node values $x_{ki}$:
\begin{equation} \label{eq: qrcy}
\hat{\bm{y}}_{t} = \sum_{i} x_{ki}w_i,     
\end{equation}
where $x_{ki}=\mathrm{Tr}\left[O_z^i \rho\left((k+v/V)\tau \right) \right]$ and $i = n+vN$ represents the order in the $v^{th}$ subinterval. The coefficients $\{w_i\}_i$ are optimized to minimize the linear regression error between the target $\bm{y}_t$ and the QRC output $\hat{\bm{y}}_{t}$, 
\begin{equation}
    \min_{\mathbf{w}} \| \mathbf{y} - \mathbf{X}\mathbf{w} \|^2,
\end{equation}
where $\mathbf{w}$, $\mathbf{y}$ and $\mathbf{X}$ are matrix forms corresponding to $w_i$, $\bm{y}_{t}$ and $x_{ki}$, respectively.The main difference between QRC and our QDE method lies in the placement of training parameters. The QDE method has parameters within the circuit, whereas the QRC model's parameters are in the linear regression coefficients, with number of $(2n_q-1)\times V+1$.

In our demonstrations, the state injection function is chosen as $\rho(s_t)=\ket{\Phi(s_t)}\bra{\Phi(s_t)}$, where $\ket{\Phi(s_t)} =\sqrt{(1-s_t)/2}\ket{0}+\sqrt{(1+s_t)/2}\ket{1}$, considering the signal is one-dimensional and rescaled to the range $[-1,1]$.
\begin{figure*}[!ht]
    \centering
    \includegraphics[width=0.8\textwidth]{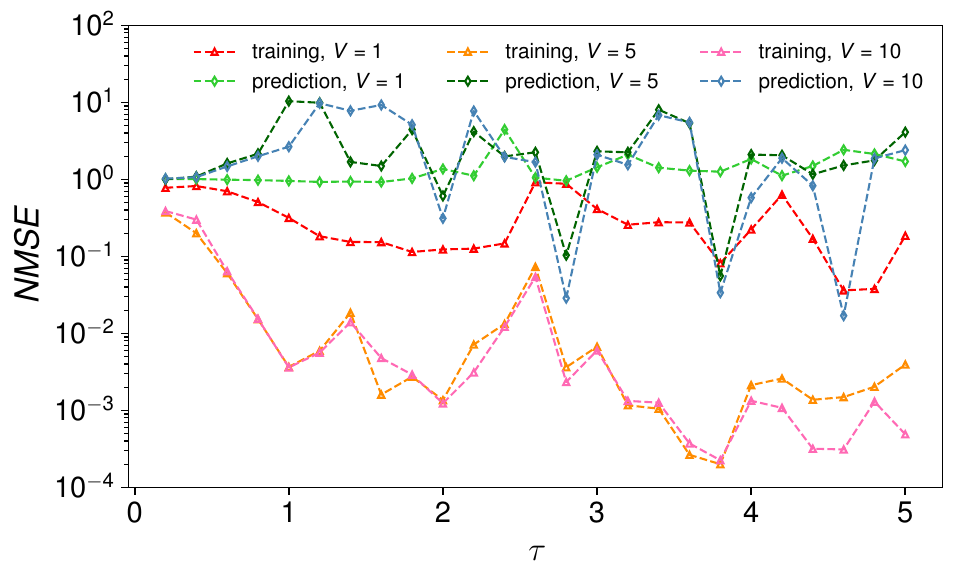}
    \caption{\textbf{Training and Predicting Error of Different Evolution Time.} 
    }
    \label{fig: qrc_tau}

\end{figure*}

\begin{figure*}[!ht]
    \centering
    \includegraphics[width=0.8\textwidth]{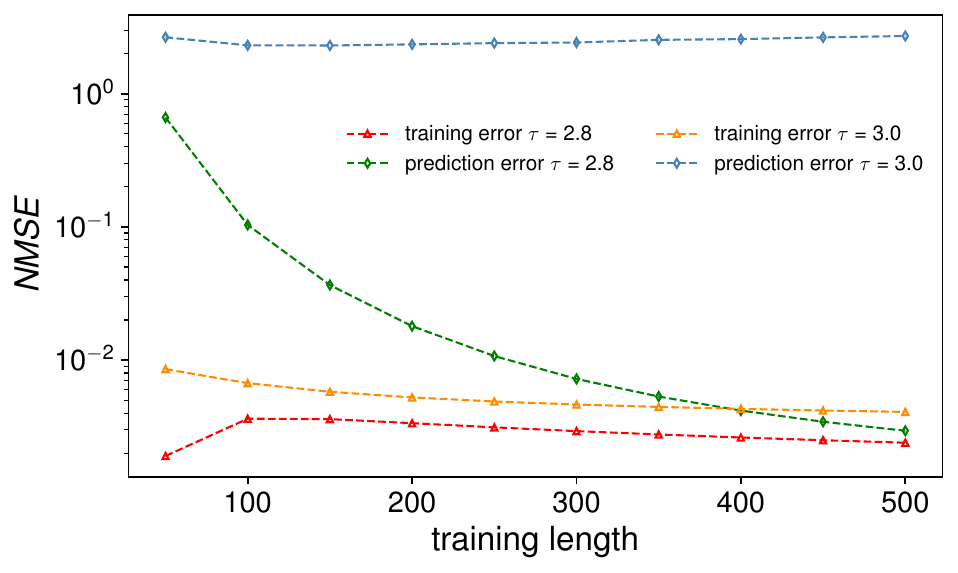}
    \caption{\textbf{Training and Predicting Error of Different Training Length.} 
    }
    \label{fig: qrc_training}

\end{figure*}

\begin{figure*}[!ht]
    \centering
    \includegraphics[width=0.8\textwidth]{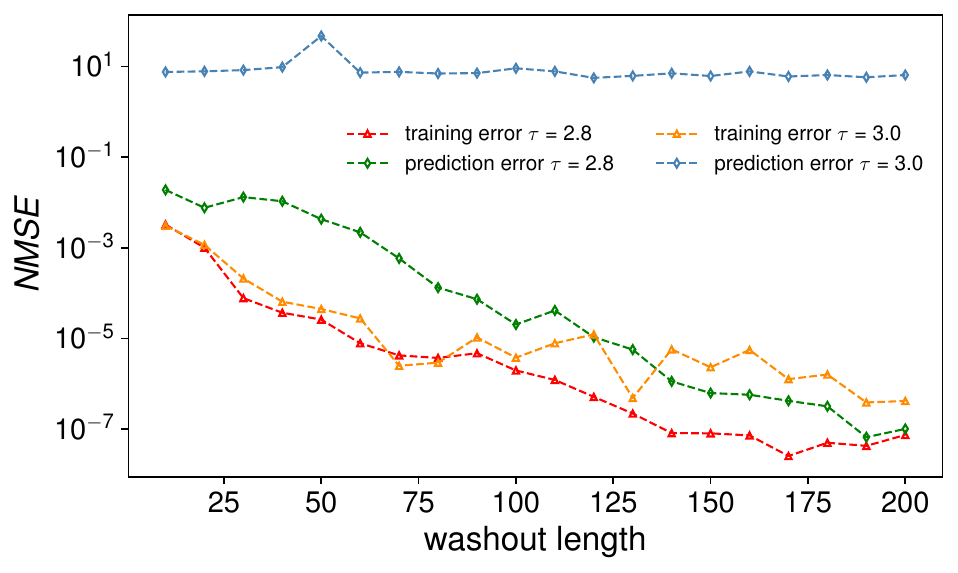}
    \caption{\textbf{Training and Predicting Error of Different Washout Time.} 
    }
    \label{fig: qrc_washout}

\end{figure*}

\begin{figure}[!th]
    \centering
    \begin{tikzpicture}
        \node[anchor=north east] (img) at (0,0) {\includegraphics[width=1.0\textwidth]{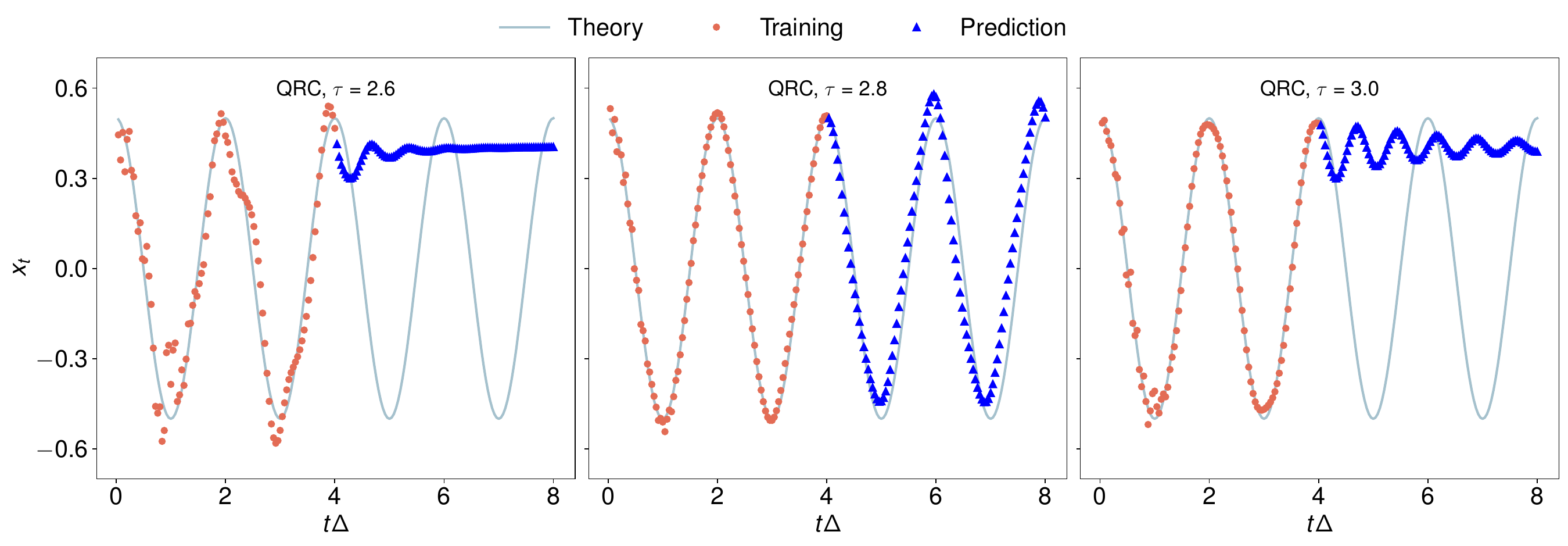}};
        
        \node[anchor=north east, inner sep=2pt] at ([xshift=-16.3cm, yshift=-0.8cm] img.north east) {\textbf{(a)}};
        \node[anchor=north east, inner sep=2pt] at ([xshift=-10.65cm, yshift=-0.8cm] img.north east) {\textbf{(b)}};
        \node[anchor=north east, inner sep=2pt] at ([xshift=-5.1cm, yshift=-0.8cm] img.north east) {\textbf{(c)}};
    \end{tikzpicture}
    \caption{\textbf{Demonstrations of QRC for Time Series Prediction}: (a), (b) and (c) are the predictions of cosine-wave from 3-qubit QRC models, varied by evolution time $\tau$. }
    \label{fig: comqrc}  
\end{figure}

The learning performance is evaluated using the normalized mean square error (NMSE). The efficacy of QRC is significantly influenced by both the evolution time $\tau$ and the number of time-multiplexing intervals $V$, as shown in Fig.~\ref{fig: qrc_tau}. 
While specific values of $\tau$, such as $\tau=2.8$, can lead to low NMSE in time series prediction, neighboring values like $\tau=2.6$ or $\tau=3.0$ may result in different performance. 
It is observed that the NMSE between predictions and targets decreases as the number of time-multiplexing intervals increases. However, this improvement becomes less pronounced beyond a certain point, such as when the number of intervals reaches $10$, indicating a limit to the method's effectiveness compared to a lower number like $5$. 

Additionally, the extension of both training length and washout period, when paired with a suitable evolution time $\tau$, can substantially enhance the QRC's learning ability. This is clearly shown in Fig.~\ref{fig: qrc_training}, Fig.~\ref{fig: qrc_washout}. In Fig.~\ref{fig: comqrc}, we plot predicting results of the QRC model on the cosine-wave signal task, which implies that these improvements are only evident with an optimal evolution value, such as $\tau=0.28$. In cases where the evolution time is not optimal, the prediction error does not decrease with increasing length. This highlights one of the significant challenges faced by QRC: the random selection of hyper-parameters in the Hamiltonian $H$ prior to training, which significantly influences the model's learning capacity.

\section{Comparision of QDE with QRC} \label{sec: compqrc}

\begin{figure}[htbp]
    \centering
    \begin{tikzpicture}
        \node[anchor=center] (a) at (0,0) {\includegraphics[width=0.7\linewidth]{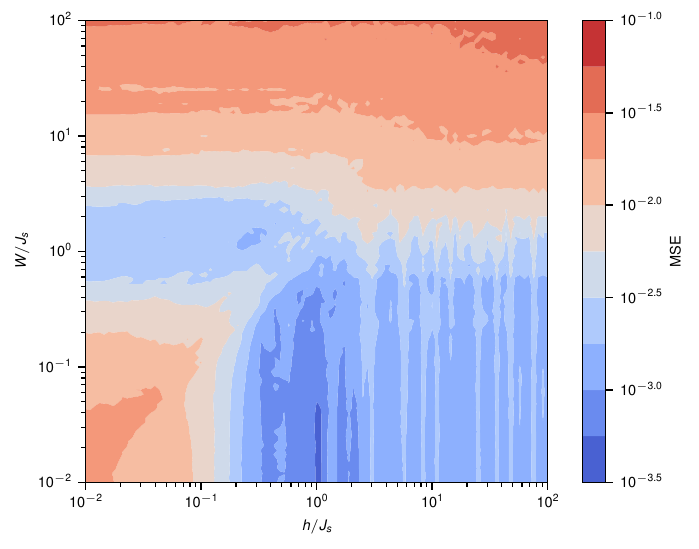}};
        
        \node[anchor=center] (b) at ($(a.south)-(0.25\linewidth,4cm)$) {\includegraphics[width=.48\linewidth]{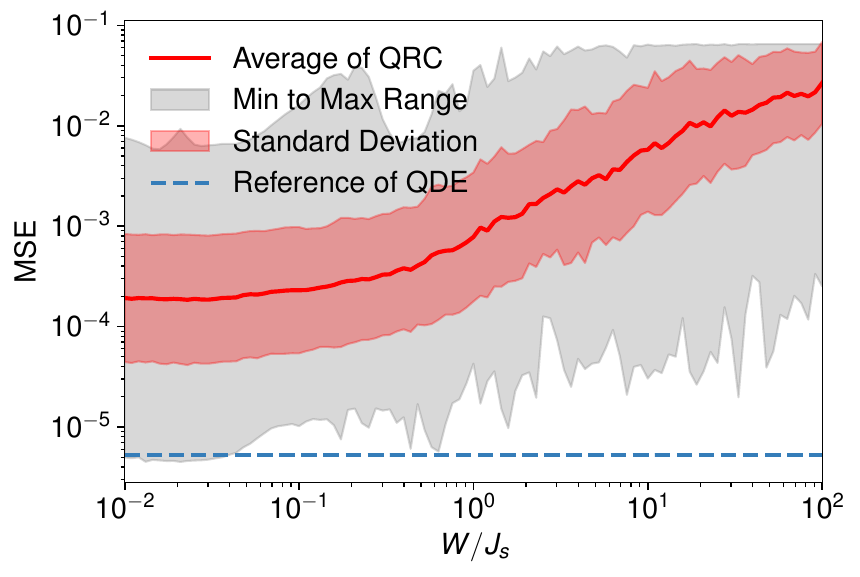}};
        
        \node[anchor=center] (c) at ($(a.south)+(0.25\linewidth,-4cm)$) {\includegraphics[width=.48\linewidth]{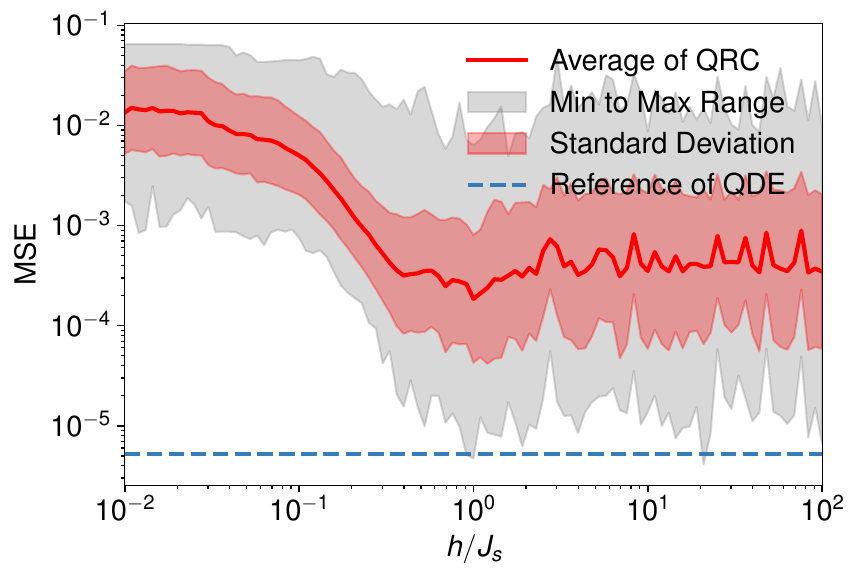}};
        
        \node[anchor=north west] at ($(a.north west)-(0.2cm,0.2cm)$) {\textbf{(a)}};
        \node[anchor=north west] at ($(b.north west)-(0.2cm,0.2cm)$) {\textbf{(b)}};
        \node[anchor=north west] at ($(c.north west)-(0.2cm,0.2cm)$) {\textbf{(c)}};
    \end{tikzpicture}
     \caption{\textbf{Comparison of QDE with QRC on Periodic Signal Prediction.} (a) Error map for different values of the Hamiltonian hyperparameters of QRC, with the transverse field $h$ and the disorder bound $W$ varying logarithmically in the range $[10^{-2}, 10^2]$. Results are averaged over 100 realizations. (b) and (c) With $h/J_s$ ($W/J_s$) fixed at the minimum error from map (a), the other parameter $W/J_s$ ($h/J_s$) is varied. The average results are shown with the minimum to maximum error range (grey shadows) and standard deviation (red shadows).}
    \label{fig: errorbar2}
\end{figure}

\begin{figure}[htbp]
    \centering
    \begin{tikzpicture}
        \node[anchor=center] (a) at (0,0) {\includegraphics[width=0.7\linewidth]{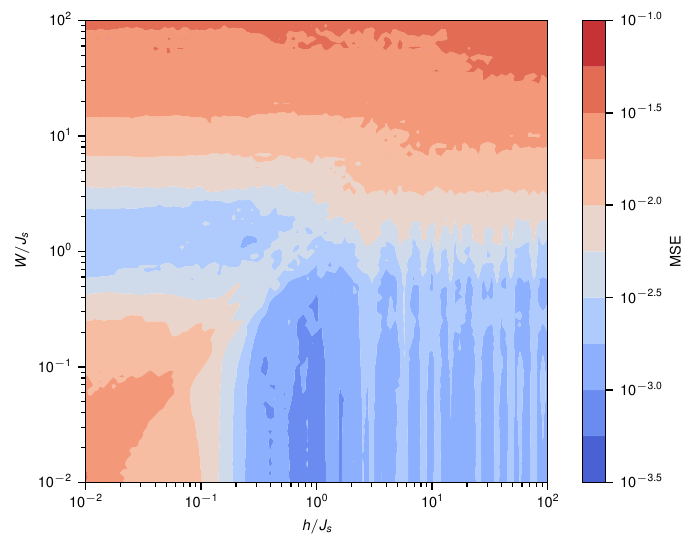}};
        
        \node[anchor=center] (b) at ($(a.south)-(0.25\linewidth,4cm)$) {\includegraphics[width=.48\linewidth]{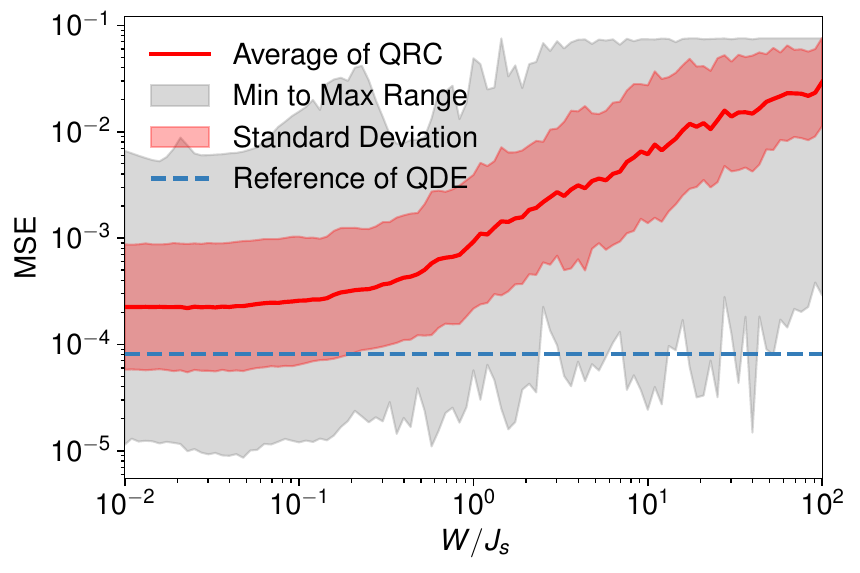}};
        
        \node[anchor=center] (c) at ($(a.south)+(0.25\linewidth,-4cm)$) {\includegraphics[width=.48\linewidth]{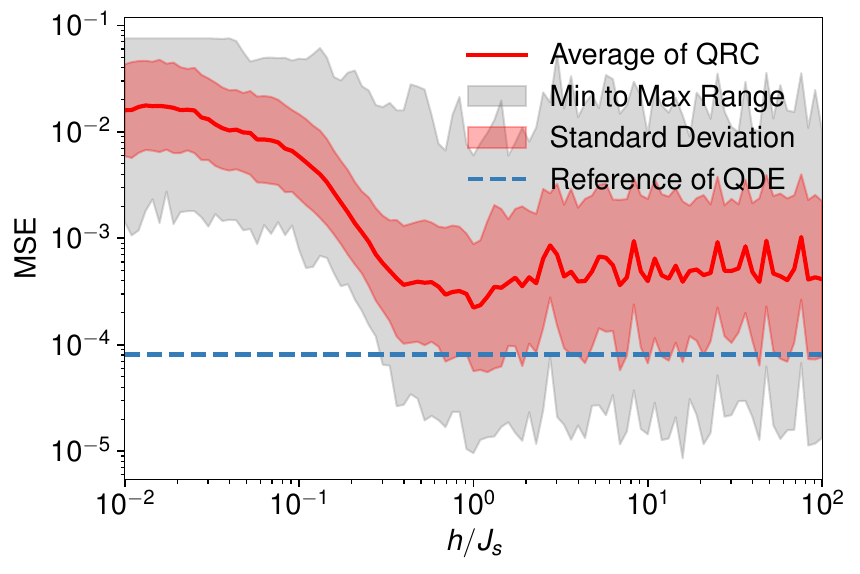}};
        
        \node[anchor=north west] at ($(a.north west)-(0.2cm,0.2cm)$) {\textbf{(a)}};
        \node[anchor=north west] at ($(b.north west)-(0.2cm,0.2cm)$) {\textbf{(b)}};
        \node[anchor=north west] at ($(c.north west)-(0.2cm,0.2cm)$) {\textbf{(c)}};
    \end{tikzpicture}
     \caption{\textbf{Comparison of QDE with QRC on Aperiodic Signal.} The error map and curves for different values of the Hamiltonian hyperparameters of QRC with the same settings as Fig.~\ref{fig: errorbar1} and Fig.~\ref{fig: errorbar2}.}
    \label{fig: errorbar3}
\end{figure}

To compare the learning capabilities of QDE and QRC, we adjusted the hyper-parameters in Eq.~\eqref{eq: Hamiltonian}, sampling each Hamiltonian's coefficients uniformly within a fixed domain. MSE values are averaged over 100 realizations with the same parameter bounds for $W$, $h$, and $J_s$. As shown in Fig.~\ref{fig: errorbar2}(a) and Fig.~\ref{fig: errorbar3}(a), the MSE values vary with different transverse field $h$ and disorder bound $W$ in units of $J_s$. The minimum points of the error map for the three tasks—cosine-wave signal, periodic signal, and aperiodic signal—are $(h=1.0000, W=0.0479)$, $(h=1.0000, W=0.0302)$, and $(h=1.0000, W=0.0275)$, respectively. We also plotted the error with one parameter fixed and the other parameter varying, as shown in Fig.~\ref{fig: errorbar2}(b), (c) and Fig.~\ref{fig: errorbar3}(b), (c). All these results demonstrate that our proposed method performs better than the average level of QRC.

\section{Ability for learning from non-linear dynamics}
We also demonstrate the learnability on the Rayleigh equation, a cornerstone in the study of nonlinear dynamics. We use the van der Pol form of the Rayleigh equation~\cite{Wiggins2003}:
\begin{equation}\label{eq:Rayleigh}
\ddot{x} - \varepsilon \dot{x} (1 - \delta x^2) + \omega^2 x = 0,
\end{equation}
where $x$ is the position, $\varepsilon$ and $\delta$ represent the nonlinearity and damping strength, and $\omega$ is the angular frequency.

\begin{figure}[!tb]
    \label{fig: phaseplot}
    \centering
    \includegraphics[width=0.6\linewidth]{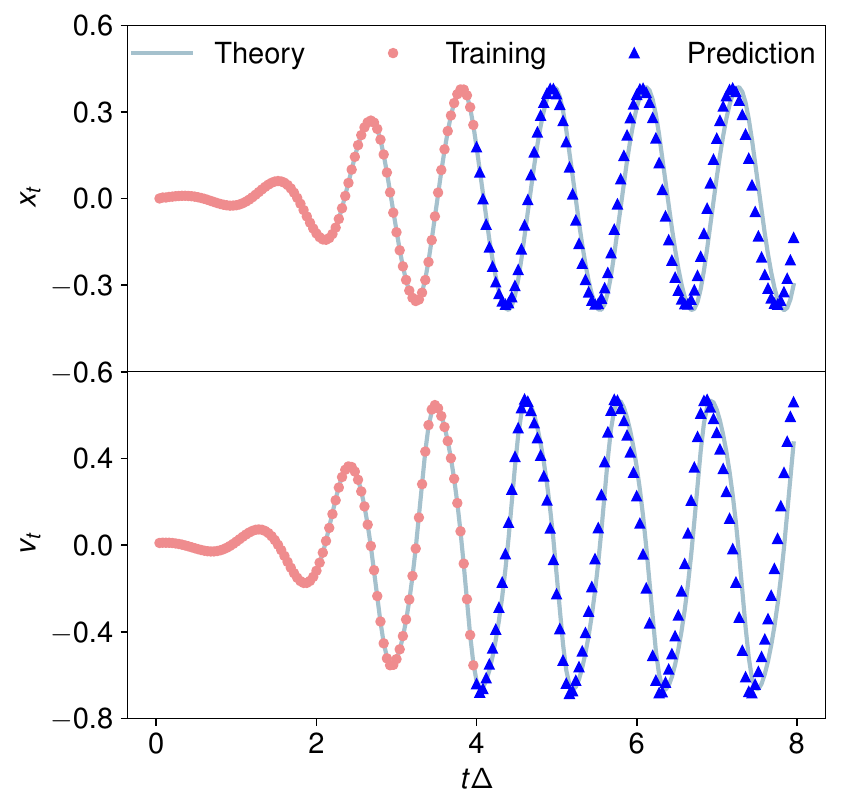}
    \caption{\textbf{Simulation of Rayleigh Dynamics Using a $(1, 2)$-QDE.} This diagram illustrates the dynamics of the Rayleigh system, focusing on displacement $x_t$ and the velocity $v_t$, which constitute the data register in the QDE. An additional qubit is allocated for the memory register. 
    }
    \label{fig: Rayleigh} 
\end{figure}

Introducing the velocity $v$, Eq.~\eqref{eq:Rayleigh} can be recast as a system:
\begin{equation}\label{eq:RayleighSystem}
\begin{cases}
\dot{x} = v, \\
\dot{v} = \varepsilon v(1 - \delta x^2) - \omega^2 x.
\end{cases}
\end{equation}

To learn the dynamics of Rayleigh equation, the QDE model employs two qubits as the data register and one qubit as the memory register. 
This demonstration utilizes the ansatz TIEA as described in Sec.~\ref{sec: settings}. 
For the Rayleigh system, we set the parameters at $\varepsilon = \pi$, $\delta = 3$, and $\omega = \pi$, initializing the system with $\vecx_0 = (0, 0.01)$. Fig.~\ref{fig: Rayleigh} presents the results, depicting two panels that represents $x(t)$ and $v(t)$ over time. 
The QDE model not only replicates the dynamics of the Rayleigh system but also exhibits consistent performance, closely following the ideal trajectory throughout the simulation.

\section{Training Procedures} \label{sec: training}
In the training stage, the circuit parameters of the QDE are optimized using the Adam optimizer within VQNet, a quantum neural network package in Python. Here We showcase the optimization process by tracking the variation of the loss function value across training epochs.

\begin{figure*}[!ht]
    \centering
    \includegraphics[width=0.8\textwidth]{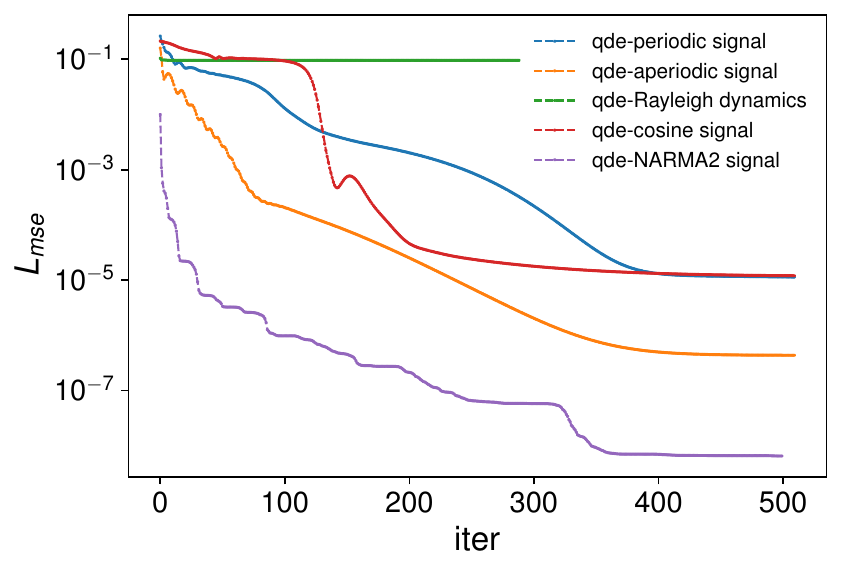}
    \caption{\textbf{Training Loss for Different Tasks} 
    }
    \label{fig: loss}

\end{figure*}
The parameters of our architecture are divided into three components. The first comprises the rotation angles of the parameterized quantum gates. The second includes the weights and biases of the learnable error-cancellation layer as shown in main text. The final component consists of the initial values of memory from different QDE blocks. 

In the simulations of a cosine-wave signal and composite signals, the QDE utilizes two linear-composite channels, each adopting the 2-qubit hardware-efficient ansatz (HEA). The simulation of the cosine-wave signal follows a similar approach but employs only one QDE channel. For the more complex Rayleigh dynamics, which requires modeling two-dimensional phase space and a one-dimensional hidden space, we use the transverse Ising evolution ansatz (TIEA).

During the training stage, the initial value $(\vecm_0,\vecx_0)$ is input into the QDE model. The autoregressively generated time series $\{\hat{ \vecx}_1, \hat{ \vecx}_2, \hat{ \vecx}_3,\dots\}$ is then collected to compute the loss function, which is defined as mean square error. In the predicting stage, the system evolves for additional $T$ steps to generate predictions. Here $T$ is set to $100$ and $1000$ for short-term and long-term predictions, respectively.

The QRC model follows the training and predicting protocol of our approach, distinct in its data injecting method. Unlike the autoregressive feeding used in the QDE model, the QRC model is trained by feeding only the target data into the Quantum Reservoir (QR) network.

\section{Quantum Chip Information} \label{sec: chip}
The experiment presented in the main text was executed on the Origin ``Wukong'' superconducting quantum computer. This platform is composed of $72$ transmon qubits arranged in a two-dimensional array, with each qubit individually controlled via an XY line and Z control. The coupling strength of these qubits is adjustable from $0$ to $80$ MHz, and the two-qubit CZ gate operates at a strength of $50$ MHz. The base frequencies of these transmon qubits are arranged in alternating high and low, the high frequency qubits have a sweetspot frequency of about $5$ GHz and the low are $4.5$ GHz.

Benchmarking results for single-qubit gates indicate an average fidelity for Pauli gates of $99.6\%$. The primary two-qubit gate of this quantum processor is the CZ gate, which achieves an average fidelity of approximately $97\%$. The gate time durations are $30$ ns for single qubit gates and $40$ ns for two-qubit gates.

Given that the experiment required only two qubits, we selected qubits Q45 and Q46 for data and memory registers of the QDE. The single gate fidelities for Q45 and Q46 are $99.59\%$ and $99.76\%$, respectively, with the CZ gate fidelity at $98.35\%$. The coherence times for these qubits are $T_1^{45}=11.8 \ {\mu}s,\ T_2^{45}=1.3 \  {\mu}s$ and $T_1^{46}=21.4 \ {\mu}s,\  T_2^{46}=0.8 \ {\mu}s$, respectively. 

Quantum circuits for the experiment were compiled using Origin Pilot and submitted through Qpandalite, a lite version of Qpanda (Quantum Programming Architecture for NISQ Device Application).

\section{Additional experiments for long-term time series prediction} \label{sec: longterm}

\begin{figure}[!th]
    \centering
    \includegraphics[width=1.0\textwidth]{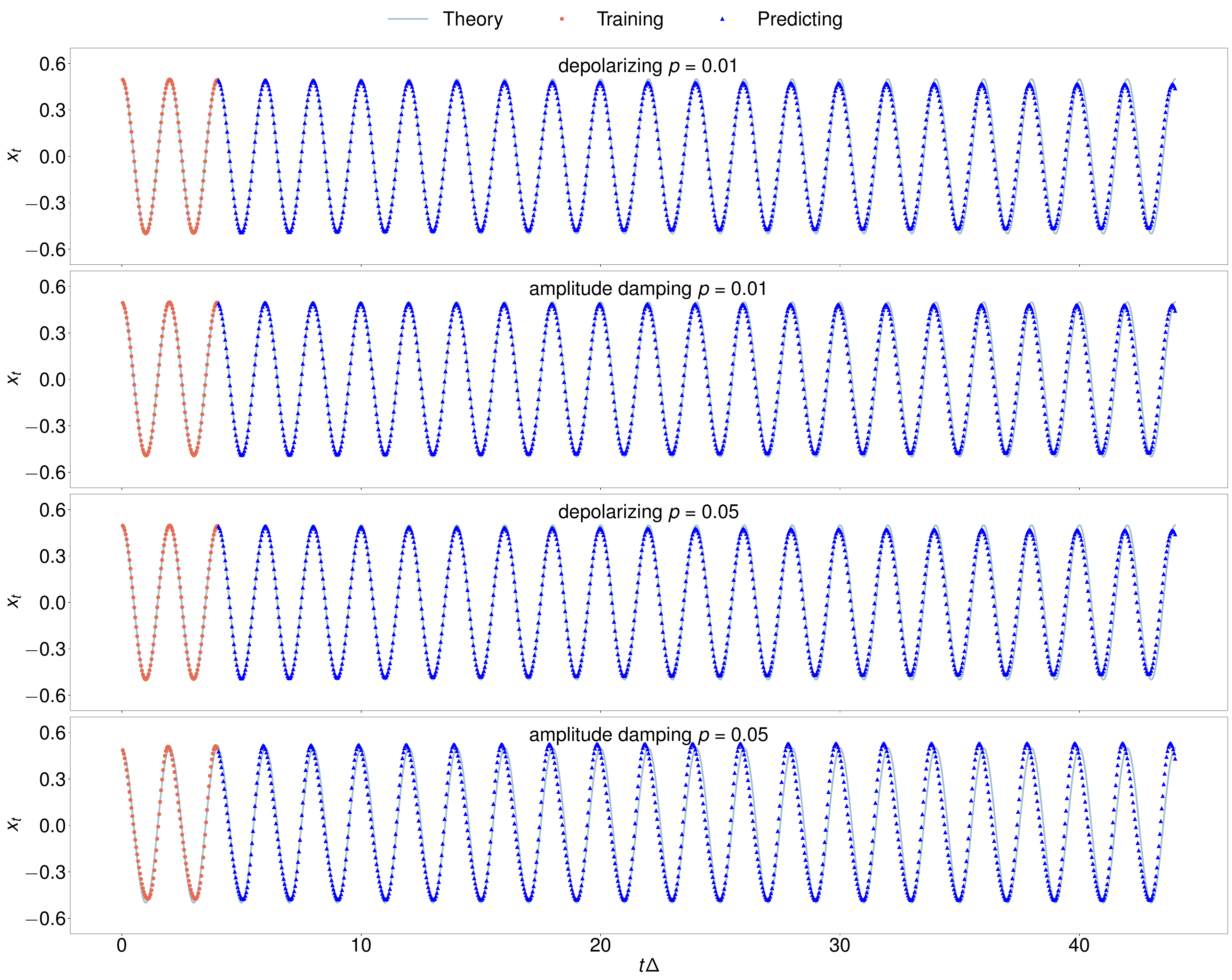}
    \caption{\textbf{Long-term Prediction on Cosine-Wave Signal with Quantum Noise.} The training and predicting protocols replicate those in Fig.~\ref{fig: applic}, with the distinction of a training-to-predicting ratio of $1:10$.  
    }
    \label{fig: longterm1}
\end{figure}

\begin{figure}[!th]
    \centering
    \includegraphics[width=1.0\textwidth]{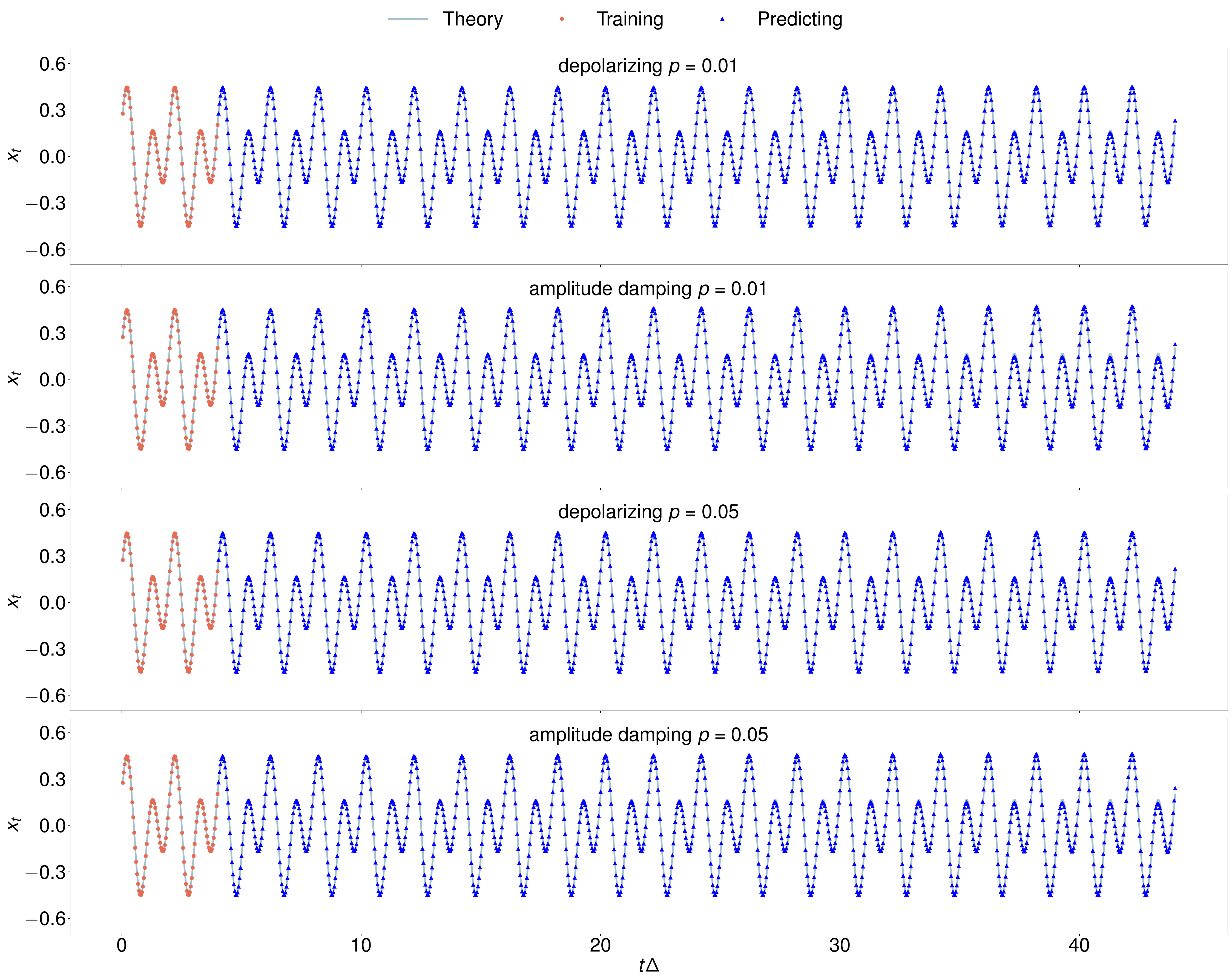}
    \caption{\textbf{Long-term Prediction on Composite Periodic Signal with Quantum Noise.} 
    }
    \label{fig: longterm2}
\end{figure}

Here, we demonstrate our results for single-mode cosine-wave signal and composite periodic signal for completeness, as shown in Fig.~\ref{fig: longterm1} and Fig.~\ref{fig: longterm2}. The circuit architecture, the training and predicting length, and the noisy model are the same in the main text of Sec.~\ref{sec: method} and Sec.~\ref{sec: results}. 
The predictions remain consistent with the theoretical curves even after 1000 steps of evolution, demonstrating the QDE model's noise resilience and its ability to capture long-term features effectively.

\begin{figure}[!th]
    \centering
    \includegraphics[width=1.0\textwidth]{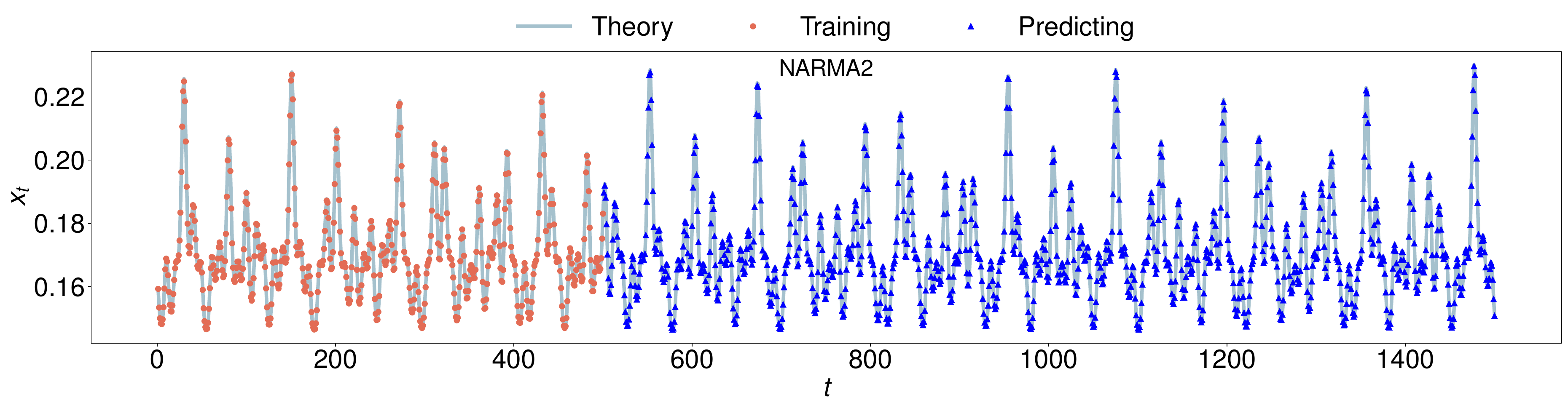}
    \caption{\textbf{Long-term Prediction on Noise-Free NARMA2 Signal.}
    }
    \label{fig: longterm4}
\end{figure}

We note that the long-term test on the NARMA2 signal is conducted in a noise-free setting as shown in Fig.~\ref{fig: longterm4}. 
Unlike the composite cosine-wave signals, which incorporate noise mitigation methods under depolarizing and phase damping noise, the NARMA2 results serve as a comparison rather than a demonstration of noise robustness. 
This choice does not imply that the method is unsuitable for noisy NARMA2 tasks.

\section{Detailed proofs}\label{sec: proofs}
Here we present detailed proofs of the Lem.~\ref{lem:1}-\ref{lem:2} and Thm.~\ref{thm:1}-\ref{thm:3} introduced in the main text.

\paragraph{Proof of Lemma 1:}
\textit{From a precise step $m(t)= \hat{m}(t)$ and $x(t)= \hat{x}(t)$, the error produced in the next step is less than $|\pi\Delta|/4$.}

\begin{proof}
For $m$, we have
\begin{equation}
\begin{aligned}
    &|m(t+1)-\hat{m}(t+1)|\\
    &= |A\sin(\pi \Delta t)[\cos(\pi \Delta)-\cos\theta] + A\cos(\pi \Delta t)[\sin(\pi\Delta)-\sqrt{1-m^2_t}\sin\theta)]|\\
    &\leq A(1-\sqrt{1-A^2})|\sin (\pi\Delta)| \\
    &< \frac{1}{4}|\pi \Delta|.
\end{aligned}
\end{equation}

By symmetry, $|x(t+1)-\hat{x}(t+1)|$ can also be bounded in the same way.
\end{proof}

\paragraph{Proofs of Lemma 2:}
\textit{
    For an erroneous step $\hat{m}(t) = m(t) + \delta_m$ and $\hat{x}(t) = x(t)  + \delta_x$, suppose $|\pi \Delta| < 1/4$, the error produce in the next step is less than $\delta+\frac{1}{4}|\pi \Delta|$ where $\delta = \max(|\delta_m|, |\delta_x|)$. }

\begin{proof}
Firstly, Consider a scenario where the error terms, $\delta_m$ and $\delta_x$ arise due to approximation. Assuming an error-free prior step, i.e., $m(t-1)=\hat{m}(t-1)$ and $x(t-1)=\hat{x}(t-1)$, it follows that $\delta_m$ and $\delta_x$ exhibit opposite signs. Without loss of generality, let $\delta_m<0$, following a similar derivation as in Lem.~\ref{lem:1}. 

Then, for $m$ we have
\begin{equation}
\begin{aligned}
    &|m(t+1)-\hat{m}(t+1)|\\
    &= |A\sin(\pi \Delta t)[\cos(\pi \Delta)-\cos\theta] +A\cos(\pi \Delta t)[\sin(\pi\Delta)-\sqrt{1-\hat{m}^2_t}\sin\theta)] +|\delta_m|\cos\theta-|\delta_x|\sqrt{1-\hat{m}_t^2}\sin\theta |\\
    &\leq A(1-\sqrt{1-A^2})|\sin (\pi\Delta)|+\delta |\cos \theta-\sqrt{1-\hat{m}_t^2}\sin\theta|.
\end{aligned}
\end{equation}
Knowing that $\pi \Delta < 1/4$, then we derive $|\cos \theta-\sqrt{1-\hat{m}_t^2}\sin\theta|_{\theta =\pi\Delta}<1$. Finally, we obtain
\begin{equation}
    |m(t+1)-\hat{m}(t+1)| < \delta+\frac{1}{4}|\pi \Delta|.
\end{equation}

By symmetry, the bound of $|x(n+1)-\hat{x}(n+1)|$ can also established analogously.
\end{proof}

\paragraph{Proof of Theorem 1:} 
\textit{
    By properly selecting $\theta_1$ and $\theta_2$, QDE can approximate $A\cos (\pi\Delta t+\phi)$, where $|A|< 1/2$, $|\pi\Delta|<1/4$ with a bounded error.
}

\begin{proof}
    Firstly, the phase $\phi$ primarily influences the initial points $(m_0,x_0)$, without contradicting the results of Lem.~\ref{lem:1} and Lem.~\ref{lem:2}. For simplicity, let's set $\phi=0$. 

    Applying Lem.~\ref{lem:1} and Lem.~\ref{lem:2} with the initial conditions $m(0)=\hat{m}(0)$ and $x(0)=\hat{x}(0)$, and setting $-\theta_1=\theta_2=\pi\Delta$, we obtain:

    \begin{equation}
    \begin{aligned}
        |m(1)-\hat{m}(1)|&<\delta+\frac{1}{4}|\pi\Delta|\\
        |m(2)-\hat{m}(2)|&<(\delta+\frac{1}{4}|\pi\Delta|)+\frac{1}{4}|\pi\Delta|\\
        |m(3)-\hat{m}(3)|&<(\delta+\frac{2}{4}|\pi\Delta|)+\frac{1}{4}|\pi\Delta|\\
        &\vdots\\
        |m(n)-\hat{m}(n)|&<\delta+\frac{n}{4}|\pi\Delta|
    \end{aligned}
    \end{equation}
By symmetry, the bound for $|x(n+1)-\hat{x}(n+1)|$ can be similarly established:
\begin{equation}
    |x(n)-\hat{x}(n)|<\delta+\frac{n}{4}|\pi\Delta|
\end{equation}
Therefore, under the worst-case scenario, QDE approximates \(A\cos (\pi\Delta t + \phi)\) with a linearly increasing error.
\end{proof}

\paragraph{Proof of Theorem 2:}
\textit{
    Given a composite signal formed by the superposition of multiple cosine-waves, $\sum_i^K A_i\cos(\omega_i t+\phi_i) $, where each term has a distinct amplitude $A_i$, frequency $\omega_i$ and phase $\phi_i$. A $K$-channel QDE can approximate this composite signal with a bounded error by properly selecting parameters $\theta_1^1,\theta_2^2,\dots , \theta_1^K,\theta_2^K$, with each superscript indexing the corresponding channel. }

\begin{proof}
    Each channel in the $K$-channel QDE contributes to the composite signal independently. According to Thm.~\ref{thm:1}, each individual cosine-wave component, represented by $A_i\cos(\omega_i t+\phi_i)$, can be approximated with a bounded error when parameters $\theta_1^i,\theta_2^i$ are optimally selected for each channel. Finally, the composite cosine-wave can be approximated with a bounded cumulative error.
\end{proof}

\paragraph{Proof of Theorem 3:}
\textit{
    If $g(t)$ is continuous on an interval $[a,b]$, then there exists a $K$-channel QDE approximating $g(t)$ with a bounded error for any $t$ in $[a,b]$. 
}
\begin{proof}
    We can choose $c>b$ as the end of new interval $[a,c]$, which can be rescaled to $[a',c']$ where $c'-a'=2\pi$. Then from Cor. 2 in Sec.1 of~\cite{tolstov2012}, there exists a trigonometric polynomial of the form $$\sigma_J(t')=\alpha_0 +\sum_{j=1}^J(\alpha_j\cos jt'+\beta_j\sin jt')$$ for which $$|g(t')-\sigma_J(t')|\leq \epsilon $$ for any $t'$ in $[a',  c']$. The polynomial $\sigma_J(t')$ can then be approximated using a $K$-channel QDE as described in Thm.~\ref{thm:2}, where in this case $K=2J+1$.
\end{proof}

\end{document}